\documentstyle[aps,pre,psfig,floats]{revtex}
\long\def\wideabs#1{\twocolumn[\hsize\textwidth\columnwidth\hsize%
\csname @twocolumnfalse\endcsname #1 \vskip1pc]}
\newcommand{\bml}{\begin{mathletters}}
\newcommand{\eml}{\end{mathletters}}
\newcommand{\beqa}{\begin{eqnarray}}
\newcommand{\eeqa}{\end{eqnarray}}
\newcommand{\ssection}{\subsection}
\newcommand{\C}{{\cal C}}
\newcommand{\G}{{\cal G}}
\newcommand{\beq}{\begin{equation}}
\newcommand{\eeq}{\end{equation}}
\begin{document}
\draft
\title{State Hierarchy Induced by Correlated Spin Domains 
in short range spin glasses.}
\author{
Eytan Domany$^{1,2}$\footnote{Electronic Address: fedomany@wicc.weizmann.ac.il},
Guy Hed$^1$, 
Matteo Palassini$^3$, and
A.~P.~Young$^4$
\\[2mm]
}
\address{$^1$ Department of Physics of Complex Systems,
Weizmann Institute of Science, Rehovot 76100, Israel\\
$^2$ Institute of Theoretical Physics, University of California Santa Barbara,
CA 93106 \\
$^3$ Department of Pharmaceutical Chemistry, University of California,
San Francisco, CA 94118 \\
$^4$ Physics Department, University of California Santa Cruz, Santa Cruz, CA
95064 
}
\date{\today}
\maketitle

\begin{abstract}
We generate equilibrium configurations for the three and four dimensional
Ising spin glass with Gaussian distributed couplings at temperatures well
below the transition temperature $T_c$.
These states are analyzed by a recently proposed method using clustering. The
analysis reveals a hierarchical state space structure. 
At each level of the
hierarchy states are labeled by the orientations  of a set of correlated
macroscopic spin domains. 
Our picture of the low temperature phase of short
range spin glasses is that of a State Hierarchy Induced by Correlated Spin
domains (SHICS).
The complexity of the low temperature phase is
manifest in the fact that the composition of such a spin domain 
(i.e. its constituent 
spins), as well as its identifying label, are defined and determined by
the ``location'' in the state hierarchy at which
it appears. 
Mapping out the
phase space structure by means of the orientations assumed by
these domains enhances our ability to investigate the
overlap distribution, which we find to be non-trivial.  Evidence is also
presented that these states may have a non-ultrametric
structure.
\end{abstract}

\section{Introduction}
\label{sec:introduction}

Whereas
equilibrium properties of infinite range~\cite{SK75} spin glasses
are completely understood within the  framework of
replica symmetry
breaking (RSB)\cite{Parisi79,Parisi80,Parisi83,RSB},
spin glasses with short range interactions
are the subject of considerable current debate and controversy.
Open  questions address
the nature of the low temperature
phases\cite{Parisi79,Parisi80,Parisi83,RSB,droplet,Huse87,Fisher87}
and their theoretical
description.
Resolution of these issues by experiments or simulations
is hindered by the  extremely long
relaxation time required for equilibration.

The most widely studied model of a short-range spin glass  is
the Edwards-Anderson model of an Ising spin glass
\begin{equation}
{\cal H} = \sum_{\langle ij\rangle}J_{ij}S_iS_j \;,
\label{eq:H}
\end{equation}
where $\langle ij\rangle$ denotes nearest neighbor sites
of a simple (hyper) cubic lattice in $D$ dimensions (we will
consider $D=3$ and $D=4$) with periodic boundary conditions, $S_i=\pm 1$,
and the couplings, $J_{ij}$, are independent random variables taken from a given
distribution.
The most commonly studied distribution, and the one we
study here, is a Gaussian distribution with zero average and  standard
deviation $J =1$. 

The high temperature phase of the model is a disordered paramagnet. As the
temperature decreases below a critical temperature $T_c$, the system (in three
or more dimensions) undergoes a transition into a frozen spin-glass phase. 
In the spin glass phase, phase space is divided into ``valleys'' which we
define as
as an ergodic subset
of the phase space, i.e. a maximal subspace 
that the system can span (or visit)
as the time 
tends to infinity.
For a finite system the definition is less
clear, but a valley is usually referred to as a part of the phase space
surrounded by free energy barriers, whose height diverges as the system size
$L \rightarrow \infty$.

This definition of ``valley'' may not be identical to the
notion of a ``pure state'' which has been
used extensively
in the literature \cite{Huse87,Fisher87,Newman98,Middleton,Palassini2d}.
and which is defined in terms of the set of correlation
functions in a fixed
{\em finite} region inside the system as $L \rightarrow \infty$
with some specified boundary conditions. 
In particular, it was recently emphasized 
\cite{KM00,HKM00,PY00,Newman01}
that a spin glass can in principle have many thermodynamically 
important valleys but just two pure states. This is realized
when 
there are many valleys with
free energies which differ by an amount of order unity, and
configurations taken from different valleys have a vanishing density of 
(relative) domain walls
as $L\to \infty$ (a {\em domain wall} is a surface separating 
a region where the two configurations
are identical from a region where they are opposite). 
In contrast, if the density of domain walls is
finite ({\em i.e.} the domain walls are
space-filling), there is a non-vanishing probability to have a domain wall
in any finite region of the system, and thus to have more than two pure
states.
In this paper we will be mainly concerned with the number and organization
of valleys, and we will not investigate whether multiple valleys
correspond to multiple pure states as defined above. In the following, by
{\em state}\/ we will always mean a microstate or spin configuration.

\ssection{RSB, droplet and TNT scenarios}
There are two traditional pictures of the spin glass phase; the droplet picture 
and RSB. According to the droplet picture of Fisher and Huse
\cite{droplet,Huse87,Fisher87}, the low energy excitations are in the form of
{\it droplets} -  compact regions with low surface tension that flip
collectively. For a droplet of size $L$ the typical (e.g. median)
free energy $F_L$ scales as $L^\theta$, where $\theta$ is a dimension
dependent exponent. Furthermore, the surface of these excitations
has a vanishing density for large $L$. Therefore, thermodynamically important 
configurations have a vanishing density of
relative domain walls, and hence a trivial 
overlap (defined below) over any finite region.
It follows that in this approach 
within any finite region 
there are only two pure states, related by spin-flip symmetry.

A parameter commonly  used to measure domain wall density
is the link overlap and its distribution. Denote a configuration 
(or state)
of an $N$-spin system
by ${\mathbf S}^\mu = (S_1^\mu,S_2^\mu,...,S_N^\mu)$.
The link overlap $q^{\rm link}_{\mu\nu}$
between two configurations ${\mathbf S}^\mu$ and ${\mathbf S}^\nu$ is defined by
\begin{equation}
q^{\rm link}_{\mu\nu} = {1\over \gamma N} \sum_{\langle ij\rangle}
S^\mu_i S^\mu_j S^\nu_i S^\nu_j \;,
\label{eq:qlink}
\end{equation}
where the sum is over pairs of neighbor sites and $\gamma N$ is the
number of bonds in the system. 
If the domain wall density
vanishes, then the distribution $P(q^{\rm link})$ of the link
overlap will be trivial: $P(q^{\rm link})=\delta(q^{\rm
link}-q_0)$. At $T=0$ one has $q_0=1$, while $q_0$ decreases for $T>0$
and becomes zero at $T_c$.

Another parameter commonly considered is the spin overlap $q_{\mu\nu}$
between configurations ${\mathbf S}^\mu$ and ${\mathbf S}^\nu$;
\begin{equation}
q_{\mu\nu} = {1\over N} \sum_{i=1}^N S^\mu_i S^\nu_i \;.
\label{eq:q}
\end{equation}
If there are only two pure states, as in the droplet model,
the {\em local}\/ overlap distribution, obtained
{\em in a finite part of an infinite system}\/,
would be trivial for all $T<T_c$, i.e.
$P(q)=0.5[\delta(q-q_{EA})+\delta(q+q_{EA})]$, where $q_{EA}$ is
the average overlap inside a pure state.
In addition, most conventional interpretations of the droplet picture
\cite{Bray86,Moore98} argue that the {\em global}\/
$P(q)$, obtained from overlaps over the {\em whole}\/ system,
would also be trivial.
This is realized if the droplets (with 
positive $\theta$) are the only relevant excitations over all
length scales. However the work of
Huse and Fisher\cite{Huse87}, and also Newman and
Stein~\cite{Newman98,Newman00}, is formulated in a sufficiently general
fashion to accommodate 
a non-trivial global $P(q)$ if this arises from multiple valleys
with non space-filling domain walls. In this situation,
one would have a trivial link overlap
distribution $P(q^{\rm link})$ in the infinite system size limit.
Even though the global $P(q)$ would be non-trivial,
the {\em local}\/ $P(q)$,
would be trivial because a
vanishing density of domain
walls means that the probability that a
domain wall goes through a fixed finite part of the infinite sample also
vanishes.

Numerical work
has, so
far, indicated a
non-trivial global $P(q)$\cite{Bhatt90,Marinari98}. For example,
Marinari et al. \cite{Marinari98}
have used parallel tempering \cite{Hukushima96,MarinariPT} to
sample 3D Ising spin glasses of sizes up to $L=16$ and for
temperatures down to $T=0.7\simeq0.74T_c$. They have found that
$P(q)$ is non-trivial, and $P(0)$ does not vanish.

In the RSB picture,
the Parisi\cite{Parisi79,Parisi80,Parisi83} theory,
which is exact for the infinite range model\cite{SK75},
is assumed to also apply to short range systems. Within the RSB solution,
both $P(q)$ and $P(q^{\rm link})$ are non-trivial for $0<T<T_c$, which 
implies that the system has many valleys and also many
pure states. RSB suggests a 
tree-like hierarchical structure for the 
pure states. At every level of the hierarchy the states are divided into sets,
so that the states in a given set are closer to each other than
to states in other sets. At the next level down the hierarchy these
sets are divided into subsets, and so on. Furthermore, according to the RSB
solution the distances  between the pure states exhibit
{\it ultrametricity} \cite{RSB}: the overlap between
any two states is determined only by the lowest level in the hierarchy,
at which they still belong to the same set.
This means that for any
triplet of pure states $\mu$, $\nu$ and $\rho$ the following relation
always holds:
\begin{equation}
q_{\mu\nu} \geq \min(q_{\mu\rho},\: q_{\nu\rho}) \;.
\end{equation}

Recently, a mixed picture
has been
proposed on the basis of numerical results of ground state
computations \cite{KM00,HKM00,PY00}, in which $P(q)$ is non-trivial
but $P(q^{\rm link} )$ is trivial (hence referred to as TNT; for {\bf T}rivial and
{\bf N}on-{\bf T}rivial). 
Houdayer, Krzakala and Martin \cite{KM00,HKM00} demonstrated the existence of
macroscopic excitations with low energy cost in 3D Ising spin glasses
of sizes up to $L=11$. This suggests that the spin overlap distribution,
$P(q)$ is non-trivial at finite temperature.
Their results also indicate that the surface of these excitations
is not space-filling, which suggests that the
link overlap distribution, $P(q^{\rm link} )$, is trivial.

Palassini and Young\cite{PY00}
studied changes to the ground state of a spin glass
when a weak perturbation is applied to the bulk of the system. They
considered short range models in three and four dimensions as well as the
infinite range SK model and the Viana-Bray model. The results for the SK and
Viana-Bray models agreed with the replica symmetry breaking picture as
expected, but the data for the short range models agreed with the TNT picture.
Effects of the type of perturbation considered in Ref.~\onlinecite{PY00}
on RSB have been investigated by Franz and Parisi\cite{Franz00b}.

Katzgraber et al. \cite{KPY00} measured directly the
distributions $P(q)$ and $P(q^{\rm link})$ at finite temperature using
parallel tempering
\cite{Hukushima96,MarinariPT} Monte Carlo,
for 3D systems of linear size $L\leq 8$ at temperature $T\geq 0.2$,
and 4D systems with $L\leq 5$ and the same temperature range.
Extrapolating their results to large sizes they found that
the variance of
$P(q^{\rm link})$ vanishes as $L\rightarrow\infty$, and the distribution
converges to delta function.
They also found the distribution
$P(q)$ to be non-trivial, as in \cite{Bhatt90,Marinari98}, so their results
also agree with the TNT picture. In the TNT scenario there are many 
valleys separated by free energy barriers, but only two pure 
states \cite{Newman00b,Newman01}.

Although several pieces of work\cite{KM00,HKM00,PY00,KPY00,Palassini99} 
supported a vanishing density of domain walls (and hence 
a fractal dimension of the domain walls, $d_s$, less than the
space dimension), a large extrapolation is involved in deducing this 
result, and Marinari and Parisi\cite{Marinari00,Marinari00b,Marinari00c}
have argued, based on their own data and a somewhat
different analysis, that actually $d_s = D$, which corresponds to RSB.

\ssection{SHICS: State Hierarchy Induced by Correlated Spin Domains}

Very recently a new method of
analysis of the structure of the low temperature phase of short
range spin glasses has been introduced~\cite{HedPRL,HedEPL}.
Evidence for a novel picture of this phase, which
is consistent with the TNT scenario, 
but inconsistent with RSB
(since there is no ultrametricity),
has been
presented~\cite{HedPRL} on the basis of a ``clustering analysis''
of the degenerate ground states of the model (\ref{eq:H}) with
$J_{ij}=\pm 1$ couplings. We denote this by ``State Hierarchy Induced by
Correlated Spin Domains'' (SHICS).

\begin{figure}[t]
\centerline{ \psfig{figure=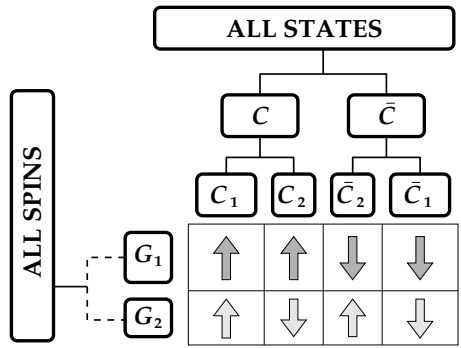,width=6cm} } \vspace{2mm}
\caption{Schematic representation of the SHICS picture; the two largest
spin domains and the first two levels in the hierarchical
organization of the states are shown. The structure of the states
is explained by the spin domains' orientations; e.g. in the states
of the two sets $\C_1,\C_2$, the spins of the larger domain,
$\G_1$, have the same orientation,  whereas the spins of the
smaller domain, $\G_2$, have flipped. Spins not in $\G_1$ or
$\G_2$ are in smaller domains which are not resolved at this level
of the hierarchy.
}
\label{fig:scheme}
\end{figure}

In
this picture there is a hierarchical tree-like structure of the
states as in the RSB solution. The highest levels of the state
hierarchy, are schematically illustrated in Fig.~\ref{fig:scheme}.
At the first level of hierarchy the states divide into sets $\C$
and $\bar\C$, such that a state in $\C$ has a counterpart with the
same energy in $\bar\C$, obtained by flipping all the spins. This
equality of the energies follows, of course, from the symmetry of
the Hamiltonian in zero field. However, this symmetry information
is not imposed on the analysis; the method finds it by itself. In
fact, it is not trivial, for a spin glass, to divide the states
into two clusters such that every state in $\C$ has its reversed
state in $\bar\C$.  Suppose, for example, that one has two states
$\mu$ and $\nu$, and states $\bar\mu$ and $\bar\nu$ with reversed
spins, such that the spin overlap $q_{\mu\nu}$ is close to zero.
Should one put $\nu$ or $\bar\nu$ in the same cluster as $\mu$?
The analysis, used in Ref.~\cite{HedPRL} and here determines which
one it is.

Many of the spins stay, with high probability, in the same
relative orientation in most of the states $\C$. Most of these
form a contiguous cluster $\G_1$, see Fig.~\ref{fig:scheme}. Among
the remaining spins, an apparently macroscopic fraction form a
contiguous domain, $\G_2$, such that the spins in it maintain,
with high probability, their relative orientation in nearly all
the states of $\C$. Hence $\C$ divides into two sub-clusters of
states, $\C_1$ and $\C_2$, depending on the orientation of $\G_2$,
see Fig.~\ref{fig:scheme}. In general, the domains $\G_1$ and
$\G_2$ are distinct. In many samples, further levels of the
hierarchy, with successively smaller domains $\G_3, \cdots$ can be
clearly resolved, as discussed later. The excitations obtained by
flipping the domains $\G_2, \G_3, \cdots$ appear to correspond to
the large scale, low energy excitations investigated by Krzakala
and Martin \cite{KM00} and Palassini and Young~\cite{PY00}.
Note that the local (or link)
overlap was
not investigated in Ref.~\onlinecite{HedPRL}. 

By contrast, in the conventional interpretations of the droplet
picture\cite{Bray86,Moore98},
the only substantial division of the states would be into
$\C$ and $\bar\C$, and any further divisions emerging from the
analysis would only correspond to microscopic spin domains. In the
RSB scenario there would be a hierarchical structure to the
states, similar to what we find here, but the nature of the spin
domains would appear to
be different, see e.g. \cite{Mezard85}. We will
discuss these differences further in Secs. \ref{sec:states} and
\ref{sec:ultm}. 

The purpose of the present paper is
to use
use the methodology of Ref.~\cite{HedPRL} to investigate whether the same
picture of the spin glass phase found there also
occurs for a spin glass with Gaussian
couplings (which has a unique ground state apart from spin reversal) at {\it
finite temperatures}. Both three and four dimensions are studied.
We find that our data do fit this picture quite well. We also present here
full details of the method.

Readers who like to skip ahead will find the picture of state clusters and
spin domains that were obtained at $T=0.2$ 
for a particular bond realization, conveniently summarized
in Fig.~\ref{fig:datmatc}. The corresponding overlap distribution $P(q)$
is presented in Fig.~\ref{fig:pq} (a).

The numerical procedure and parameters that were used in our
simulations are described in Sec. \ref{sec:numerics}. In Sec.
\ref{sec:clustering} we present the clustering methodology which
we use to identify the states hierarchy, as described in Sec.
\ref{sec:states}. In Sec. \ref{sec:spins} we use the hierarchical
partition of the state space to obtain the spin domains, show that
their sizes  scale with the system size and their correlation does
not approach unity as $L \rightarrow \infty$. We also show that these
spin domains, that were identified on physical grounds, can also
be obtained by a cluster analysis of the $N$ spins.
Those domains yield a non-trivial overlap distribution $P(q)$
with peaks corresponding to the different domain sizes, as we show in
Sec. \ref{sec:overlap}. Since we find that the average correlation
between spins in different
domains does not approach unity with increasing
system size, $P(q)$ will remain
non-trivial as $L \rightarrow \infty$.
The nature of our picture appears to yield a
non-ultrametric structure, 
as indicated at the end of Sec. \ref{sec:states}
and demonstrated
in Sec. \ref{sec:ultm}, in which we present a parameter for ultrametricity and
measure its distribution.
Finally, our method and findings are summarized in Sec. \ref{sec:summary}.

After this work
was completed we received a preprint from Marinari et al~\cite{MMZ01} who have
adopted and adapted the methodology of~\cite{HedPRL,HedEPL} to study the $J_{ij}=\pm
1$ model in $d=3$
at a single  temperature
($T=0.5$ - whereas here we considered $T=0.2$ and $T=0.5$
for the Gaussian model).
They also confirmed that the previously observed SHICS scenario~\cite{HedPRL}
of a tree-like structure of the states, 
governed by correlated spin domains,
remains valid at a non-zero temperature.

\section{Numerical Method}
\label{sec:numerics}
We simulate the Hamiltonian in Eq.~(\ref{eq:H})
using the parallel
tempering Monte Carlo method\cite{Hukushima96,MarinariPT}.
In this technique, one simulates several identical replicas of the system
at different temperatures, and, in addition to the usual local moves, one
performs global moves in which the temperatures of two replicas (with
adjacent temperatures) are exchanged. This greatly speeds up equilibration
at low temperatures.
The detailed balance condition for temperature exchanges
is satisfied by accepting
these moves with probability
$\min\left[ \exp(\Delta E \Delta \beta), \ 1\right]$, where $\Delta E = E^\mu -
E^\nu$, $E^\mu$ and $E^\nu$ are the (total) energies of
replicas $\mu$ and $\nu$, and $\Delta \beta = \beta^\mu - \beta^\nu$ is the
difference in inverse temperatures. 

We choose a set of temperatures $T_i, i = 1, 2, \cdots , N_T$, in order that
the acceptance ratio for the global moves is satisfactory, typically greater
than about $0.3$.
We use the test for equilibration
discussed in Ref.~\onlinecite{KPY00}, which involves measurements of
$q_{\rm link}$. For that, we need, at each temperature, two copies of the
system, so we actually run 2 sets of $N_T$ replicas and perform the global
moves independently in each of these two sets.

For the three-dimensional model we stored configurations
for sizes $L=4,5,6$ and $8$ at $T=0.20,
0.50$ and $2.0$, which are to be compared with \cite{Marinari98}
$T_c \simeq 0.95$. We also
stored size $L=12$ configurations at $T=0.50$.
The parameters of the
simulations are shown in Table~\ref{table:34d}. The highest temperature was
$2.0$ and lowest $0.2$ except for $L=12$ where the lowest temperature
was $0.5$.

We generated randomly chosen
interactions, $J_{ij}$, with a Gaussian  distribution with zero mean and
standard deviation unity.
For each size, temperature and bond configuration (sample) we saved $500$
spin configurations. These, together with the 500 obtained from them by spin
reversal, constitute our ensemble of $M=1000$ spin configurations, generated for
each sample.

\begin{table}
\begin{center}
\begin{tabular}{rrrrrr}
D  &  L  &  $ n_{\rm equil}$  & $n_{\rm meas} $ & $N_{\rm samp}$ & $N_T$  \\
\hline
3  & 4 &  $10^4$          & $10^5$           & 500  &   11   \\
   & 5 &  $5 \times 10^4$ &  $5 \times 10^5$ & 500  &   15   \\
   & 6 &  $3 \times 10^5$ &  $3 \times 10^6$ & 500  &   15   \\
   & 8 &  $10^6$          & $10^7$           & 335  &   18   \\
   & 12 &  $2 \times 10^5$ & $2 \times 10^6$  & 254  &   20   \\ \hline
4  &  3 & $10^4$          & $10^5$           & 500  &   13   \\
   &  4 & $4 \times 10^4$ & $4 \times 10^5$  & 500  &   13   \\
   &  5 & $8 \times 10^5$ & $8 \times 10^6$  & 200  &   25   \\ 
\end{tabular}
\end{center}
\caption{Parameters of the simulations in $D=3$ and 4 dimensions.
$N_{\rm samp}$ is the
number of samples (i.e. sets of bonds),
$n_{\rm equil}$ is the number of sweeps for equilibration and $n_{\rm meas}$
is the number of sweeps for measurements
for each of the $2 N_T$ replicas for a single sample.
$N_T$ is the number of
temperatures used in the parallel tempering method.
}
\label{table:34d}
\end{table}

For the four-dimensional
model we stored configurations for
sizes $L=3,4$ and $5$ at $T=0.2, 0.8$ and $2.6$,
compared with \cite{Parisi96} $T_c \simeq 1.80 $.  The highest temperature was
$2.6$ and the lowest $0.2$. $500$ spin configurations were saved for each 
sample. The other parameters of the
simulations are also shown in Table~\ref{table:34d}.


We are confident, based on the equilibration test used~\cite{KPY00},
that the spin configurations we generate are in thermal equilibrium.  However,
it is interesting to ask whether there are significant correlations between
them. Our results do not require that correlations be absent, but the
clustering method does require that a substantial number of independent
configurations are generated for each sample.

For each set of bonds (and temperature) we store
500 spin configurations, 250 for each replica, so the number of
sweeps between measurements, $t_{\rm meas}$, is given by $t_{\rm meas} =
n_{\rm meas} / 250$ where $n_{\rm meas}$ is given in Table \ref{table:34d} 
We will denote by ``time", $t$,
the number of Monte Carlo sweeps.
A quantity which tests
for correlations is the time-dependent Edwards-Anderson
order parameter $[q(t)]_J \equiv [ \langle S_i(t_0) S_i(t_0 + t) \rangle ]_J $,
where $\langle \cdots \rangle$ indicates a thermal average.
This is estimated from our spin configurations according to
\begin{equation}
[q(t)]_J = \left[ {1 \over N_{t_0}} \sum_{t_o}^{N_{t_0}}
{1 \over N} \sum_{i=1}^N S_i(t_0) S_i(t_0 + t) 
\right]_J \ ,
\label{eq:qt}
\end{equation}
where we have averaged over $N_{t_0}$ values for the initial 
time $t_0$ as well as over spins and bond configurations. 
Clearly $[q(0)]_J = 1$ and $[q(t)]_J \to 0$ for times sufficiently long that
there are no correlations.

\begin{figure}[t]
\centerline{\psfig{figure=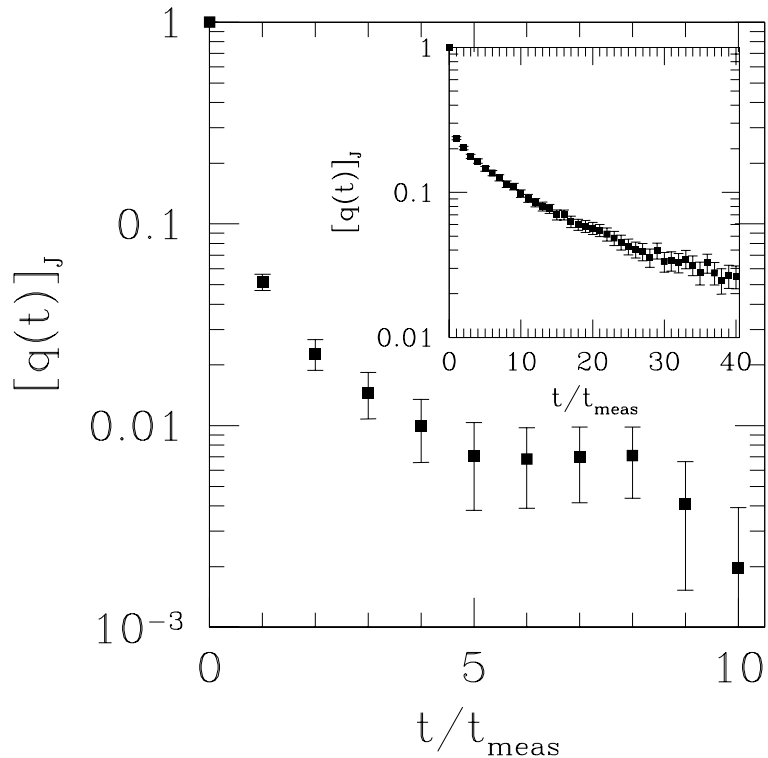,width=11cm}}
\caption{
The main part of the figure shows the
correlation between spin configurations, $[q(t)]_J$, defined in
Eq.~(\ref{eq:qt})
of the text, in $D=3$ for $L=8, T=0.2$. The horizontal axis represents the
number of Monte Carlo sweeps between the two configurations in units of the
number of sweeps between individual measurements, $t_{\rm meas}$. For
comparison, for each set
of spins (``replica''), a total of 250 configurations are generated. 
The inset shows results for $D=3, L=12, T=0.5$, which indicate that 
correlations between spin configurations are significantly larger than for
$L=8$.
}
\label{fig:qt}
\end{figure}

In Fig.~\ref{fig:qt} we show data for $[q(t)]_J$ in $D=3$
for $L=8, T=0.2$. We see that
the correlation is very small even for $t/t_{\rm meas} = 1$ (i.e. between the
configurations of neighboring measurements). The same is true for smaller
sizes and higher temperatures. For $L=12, T=0.5$, shown in the inset to
Fig.~\ref{fig:qt}, the correlations are larger, about 
0.24 for $t/t_{\rm meas} = 1$, and then decrease, though less fast than
exponentially. Thus, for $L=12$, correlations will decrease, somewhat, the
effective number of independent spin configurations.
However, we 
feel that this is not crucial since we do not use the $D=3, L=12$ data for the
clustering analysis, and only present it in one place, Figure \ref{fig:gg3d}.

In $D=4$, for $L=3$ and 5, the strength of the correlations at $T=0.2$
is small, comparable to, or
less than that for $D=3, L=8, T=0.2$. For $L=4$, the correlation is
intermediate between the results shown in $D=3$ for $L=8$ and 12. 

\section{Clustering methodology}
\label{sec:clustering}

Clustering is an important technique  to perform exploratory data analysis.
The aim is to partition data according to natural
classes present in it.  By ``natural classes" we mean groups of points that 
are close to one another and relatively far from other points, so that it
is natural to
assign them together, 
without using any preconceived 
information on the features according to which the set
should be classified. 

The standard definition of the clustering problem
\cite{Jain88} is as follows. Partition $N$ 
given data points (or objects) into $K$ groups (i.e. clusters) so that two points that
belong to the same group are, in some sense, more similar than two that
belong to different groups. The $i=1,2,...N$ data points are 
specified either in terms
of their coordinates ${\vec X}_i$
in a $D$-dimensional space (representing the measured values of 
$D$ attributes or features) or, alternatively, by means
of an $N\times N$ \ ''distance matrix'', whose elements $d_{ij}$ measure the
dissimilarity
of data points $i$
and $j$. The traditional tasks of clustering algorithms are to determine $K$
and to assign each data point to a cluster.


In the context of the present work we can think of our sample of 
$M$ spin configurations
as the objects to be clustered.
Each object is represented by an $N-$component vector
${\mathbf S}^\mu = (S^\mu_1,S^\mu_2,...,S^\mu_N)$,
where $S^\mu_i = \pm 1$ is the value 
taken by spin $i$ in state $\mu$.
An alternative view, which we also use,
is to consider the $N$ spins as the objects to be clustered.

Our first aim in this work was to 
look for a hierarchical structure of the states of a spin glass. 
Hence we wanted to
find a hierarchy of partitions, where each partition is a refinement of the
previous partition. This purpose calls for using a hierarchical clustering
algorithm. The output of such an algorithm is a tree of clusters,
called a {\em dendrogram}.  
Each node in the tree corresponds to a cluster. The splitting of a cluster
represents its partition into sub-clusters. The trunk is the single 
``cluster" that
contains {\it all} the
objects, representing the crudest partition; at the other extreme each leaf is a
cluster of a single object, representing the finest partition.

There are many  clustering algorithms that produce such
a hierarchical partitioning of any data set.
We tried two algorithms; a recently 
introduced one, SPC \cite{Blatt97}, which uses the physics of granular ferromagnets to 
identify clusters, and a graph-based algorithm proposed by Ward.
In the present problem
the
state clusters are nearly always compact (i.e. consist of a high density
of points concentrated in a relatively small volume),
and the same holds for spin clusters.
Therefore  an
algorithm that identifies compact clusters easily is most suitable for our
needs and
Ward's algorithm is designed to find such clusters. 
Furthermore, SPC is a
``short-range'' algorithm \cite{HedThesis}, 
in the sense that it couples directly only points within a 
characteristic length scale. If this scale is tuned by the distances inside
valleys, which are much smaller than the distance between them, SPC 
identifies the valleys as different clusters, but may miss their relative
hierarchical structure.

Ward's algorithm~\cite{Jain88} is
{\em agglomerative}, 
works its way up from
the leaves to the trunk, by fusing two clusters at each step. 
It begins with an initial partition to $i=1,2,...,N$ clusters, 
with a single data point in each. One calculates the distance $D_{ij}$ between
every pair of points $i,j$; one may use, for example, 
the Euclidean definition of distance, 
or (for binary valued coordinates) the Hamming distance.

At each step that pair of clusters, $\alpha,\beta$, which are 
separated by the shortest effective distance $\rho_{\alpha \beta}$
from each other, are identified and fused
to form a new cluster $\alpha^\prime=\alpha\cup\beta$. 
The process stops when there is only one
cluster, that contains all points.

Initially each data point $i=1,2,...N$ constitutes a cluster and hence 
the distance $\rho_{ij}$ between two such ``clusters" is the original
distance $D_{ij}$ between points $i$ and $j$. For subsequent steps, however,
one must define an effective distance
$\rho_{\alpha \beta}$, between any two clusters $\alpha$ and $\beta$. 
This distance is defined by the following update rule:
if at a particular step we fuse two clusters, $\alpha$ and
$\beta$, to form a new cluster $\alpha^\prime$, 
we calculate the  effective
distances $\rho_{\gamma \alpha^\prime}^\prime$, between every
unchanged cluster, $\gamma \neq \alpha,\beta$,
and the new $\alpha^\prime$, according to 
\beq
\rho^\prime_{\alpha^\prime \gamma}=
{n_\alpha+n_\gamma\over n_\alpha+n_\beta+n_\gamma}\rho_{\alpha\gamma} +
{n_\beta+n_\gamma\over n_\alpha+n_\beta+n_\gamma}\rho_{\beta\gamma} -
{n_\gamma\over n_\alpha+n_\beta+n_\gamma}\rho_{\alpha\beta} \;,
\label{eq:Ward}
\eeq
where $n_x$ is the number of data points in cluster $x$.	
Distances between unfused clusters
remain the same.
Note that $\rho^\prime_{\alpha^\prime\gamma}>\rho_{\alpha\beta}$ and
$\rho^\prime_{\gamma\delta}>\rho_{\alpha\beta}$ for every two clusters
$\gamma, \delta$. Hence after every fusion step the minimal distance between 
clusters increases.

Whenever two clusters are fused, the quantity
\beq
S = \sum_\alpha \sigma_\alpha
\eeq
where $\sigma_\alpha$ is
the sum of squared distances over all pairs of points in cluster $\alpha$,
\beq
\sigma_\alpha = \sum_{i,j \in\alpha}{D_{ij}}^2 .
\eeq 
increases.  
It can be shown~\cite{Jain88} that Ward's fusion and  distance update rules
ensure that at each fusion step this increase is minimal.

We associate a value $\tau$ with each cluster $\alpha^\prime$, where 
$\tau(\alpha^\prime)=\rho_{\alpha\beta}$ is
the effective distance between the two clusters that
were fused to form $\alpha^\prime$.	
For the initial single-point clusters we set $\tau=0$.
$\tau(\alpha)$
is related to $\sigma_\alpha$, the sum of squared distances within cluster $\alpha$. 
Clusters formed earlier have
lower $\tau$ values, and their $\sigma_\alpha$ is smaller. 

\begin{figure}[t]
\centerline{\psfig{figure=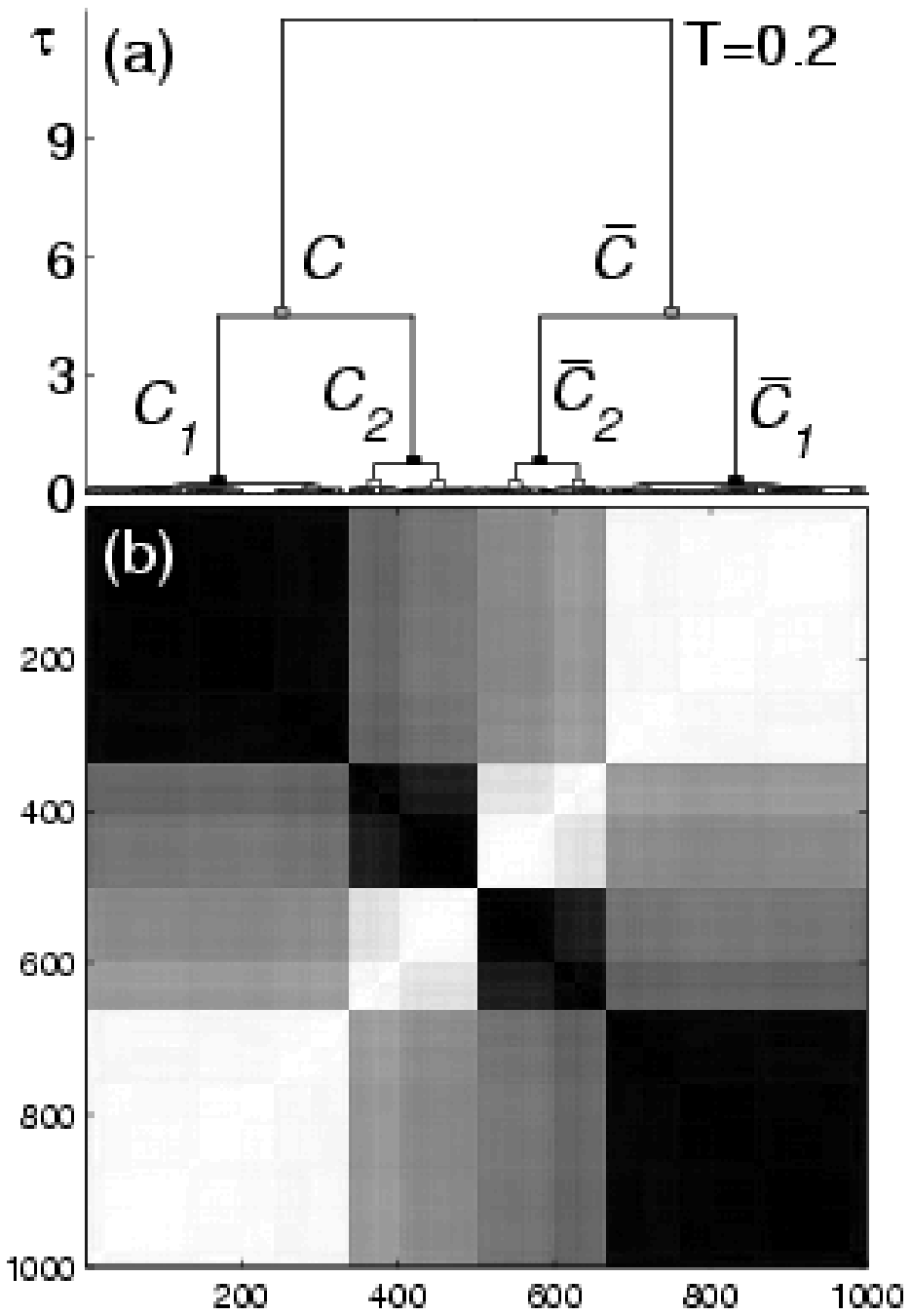,width=6cm}
\psfig{figure=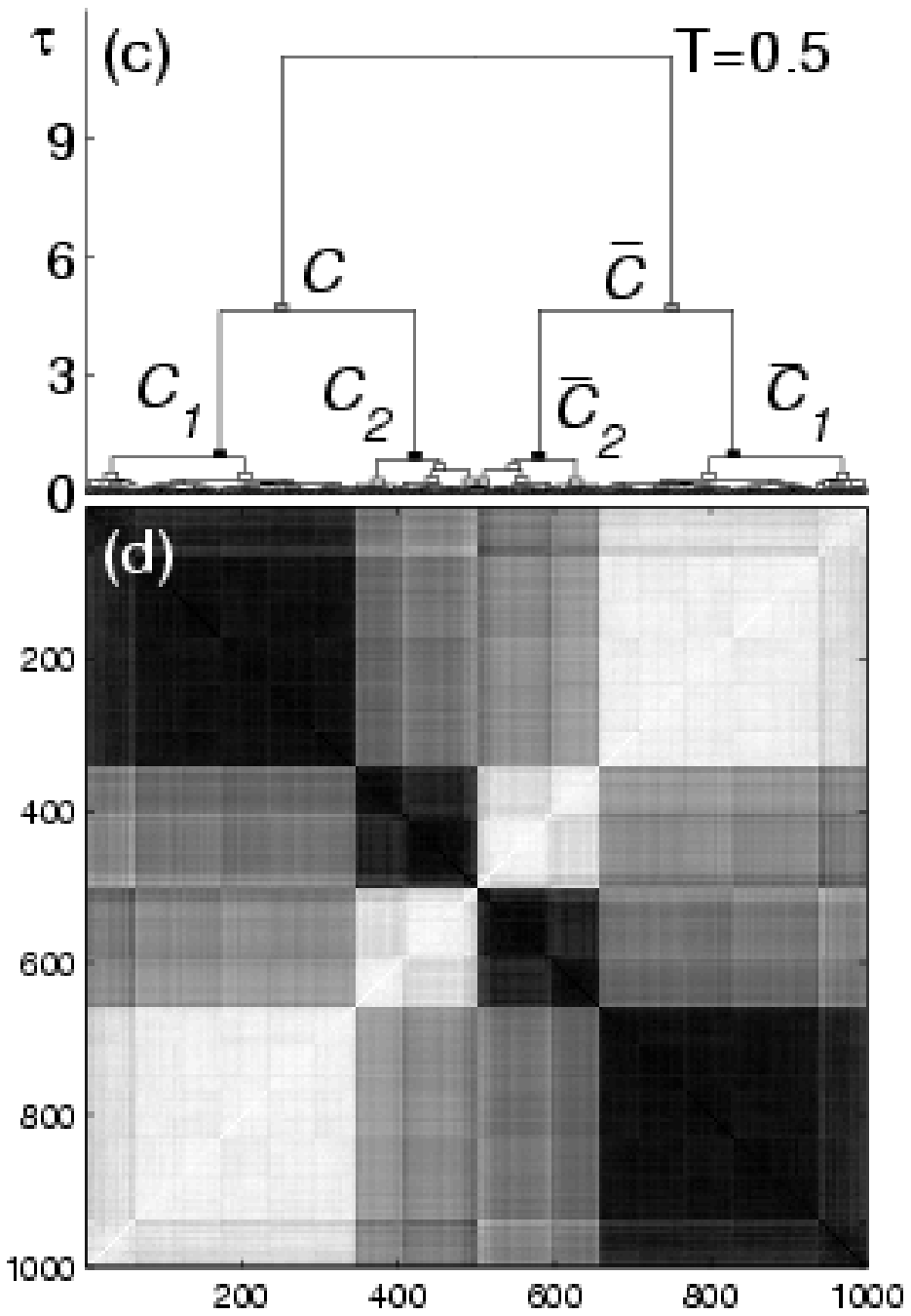,width=6cm}
\psfig{figure=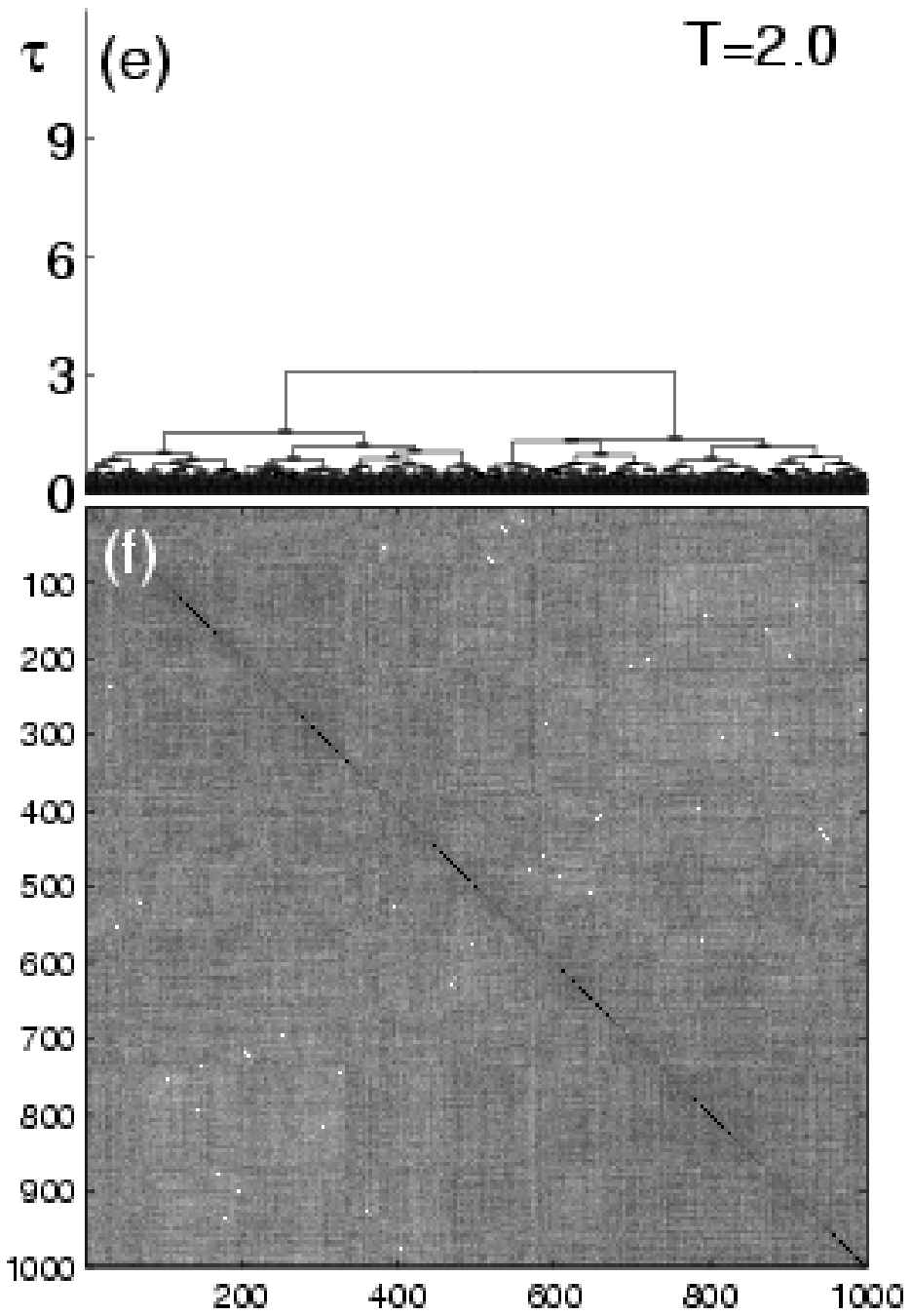,width=6cm}}
\vspace{2mm}
\caption{
{\bf (a)} The dendrogram obtained by clustering the
$M=500\times2$ states of a specific
realization in $D=3$ of size $N=8^3$ at $T=0.2$. The vertical axis describes
the
value of $\tau$, defined in Section \ref{sec:clustering}.
{\bf (b)} The distance matrix of the states used as an input to Ward's
algorithm. Darker shades correspond to smaller distances.
The states are ordered according to their position
on the dendrogram (a).
{\bf (c), (d)} The same as in (a), (b), for the same realization $\{ J \}$,
but for an ensemble of states obtained at at $T=0.5$.
{\bf (e), (f)} The same as in (a), (b), for the same realization, at $T=2.0$,
which is greater than $T_c \simeq 0.95$.
Note that this dendrogram is not symmetric; almost all the distances are
close to
0.5, so at each stage of the algorithm there were several possible partitions
that gave minimal value to $S$. In the implementation we used, the
algorithm chose a non-symmetric partition.
}
\label{fig:state_dend}
\end{figure}

The result of the algorithm is a dendrogram, or tree, as in Fig.
\ref{fig:state_dend}(a). 	
The leaves at the bottom represent the individual data points; 
they are ordered on the
horizontal axis in a way that reflects their proximity and hierarchical
assignment~\cite{Alon99}.
The small boxes at the nodes represent clusters. The vertical location of
cluster $\alpha$ is its $\tau$ value, and is thus related to its $\sigma$.
When two relatively tight and well-separated
clusters are fused, the $\tau$ value of the resulting cluster is	
much higher than those of the two constituents. Hence the length
of the branch {\it above} cluster $\alpha$ provides a measure of its relative
$\sigma_\alpha$;  long branches identify clear, tight  clusters.


Every clustering algorithm is designed to work well for data that satisfy some
(usually implicit) assumptions. When the actual distribution of the data points
deviates from these assumptions, the algorithm may produce some ``unnatural" 
partitions. For Ward's algorithm one has to look out for two potential problems.

{\it   The first problem} arises from the implicit assumption that 
minimizing $S$, the variance within clusters, leads to ``natural" partitions.
This is not the case when, for example, the data consists of
a set of points $C$ whose natural partition is into 
two clusters $C_1$ and $C_2$ with very different sizes. 
We encountered this problem  only 
for the classification of very small groups of states,
and therefore it has very little statistical effect on our results.

{\it The second,} and seemingly more serious concern is the fact that like 
every agglomerative
algorithm,
Ward's algorithm will generate a tree-like structure when applied to 
{\it any} set of
data. 
In fact, it is fairly easy to identify
when the dendrogram and the corresponding partitions do correspond to
real hierarchical structure, and when is it an artifact of the clustering
algorithm used. We used three indicators for 
the ``naturalness" of our state clusters: direct observations of 
(1) the dendrograms
and (2) the distance matrices, as well as (3) a quantitative measurement 
of the sizes of our
clusters, which are significantly smaller than the distance between clusters.
These points are demonstrated in Sec. \ref{sec:states}; 
for a detailed discussion see ~\cite{HedThesis}.

\section{State space structure}
\label{sec:states}

For a particular (randomly chosen) set of bonds $\{ J \}$ of the system
we generate, as discussed in Sec.~\ref{sec:numerics},
a sample of $500$ states, which constitute an equilibrium ensemble at
a temperature $T$. Next, we add to this ensemble the set of $500$ states
obtained from the original set by spin reversal. Clearly the new ensemble of
$M=1000$
states also corresponds to thermal equilibrium \cite{foot2}
at $T$.
We now address the following question:
\begin{quote}
Do the $M$ states of the equilibrium
ensemble cover the $2^N$ points of state space or a part of it uniformly,
or is there some underlying hierarchical organization?
\end{quote}
As it turns out, the answer depends on
$T$; whereas above $T_c$ the $M$ states do not exhibit any apparent
structure, below $T_c$ a very pronounced hierarchical organization is
seen. To uncover this organization we use the clustering methodology of
the previous Section, treating the $M$ states of our ensemble as the data
points to be clustered.

\begin{figure}[t]
\centerline{\psfig{figure=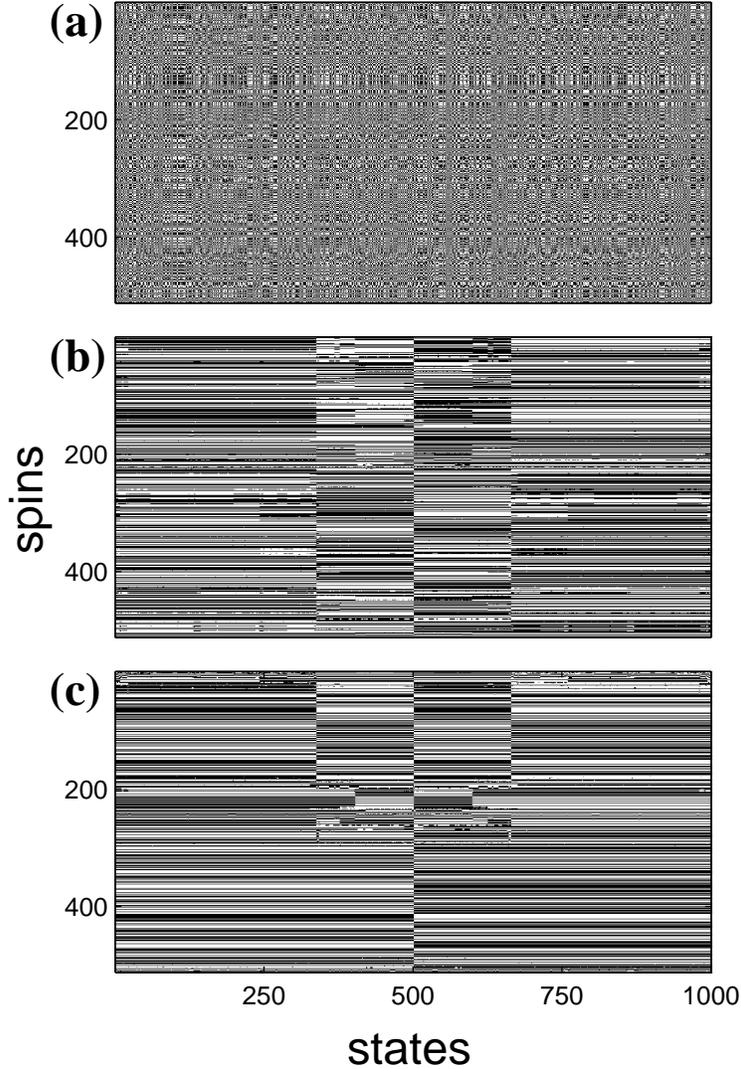,width=10cm}}
\vspace{2mm}
\caption{
{\bf (a)} The original data matrix of $500\times2$ states $\mathbf S^\mu$,
$S^\mu_i=\pm1$, with black/white representing $+/-$.
This $3 D$ sample was generated for a
realization of size $8^3$ at $T=0.2$ (the same
one as in Fig. \ref{fig:state_dend}).
The spins are in lexicographic order.
{\bf (b)} The same matrix, with the states ordered according to the
dendrogram in Fig. \ref{fig:state_dend}.
{\bf (c)} The matrix in (b) with, in addition, the spins ordered according
to the spin dendrogram $\cal D$ in Fig. \ref{fig:spin_dend}.
}
\label{fig:datmat}
\end{figure}

\begin{figure}[t]
\centerline{\psfig{figure=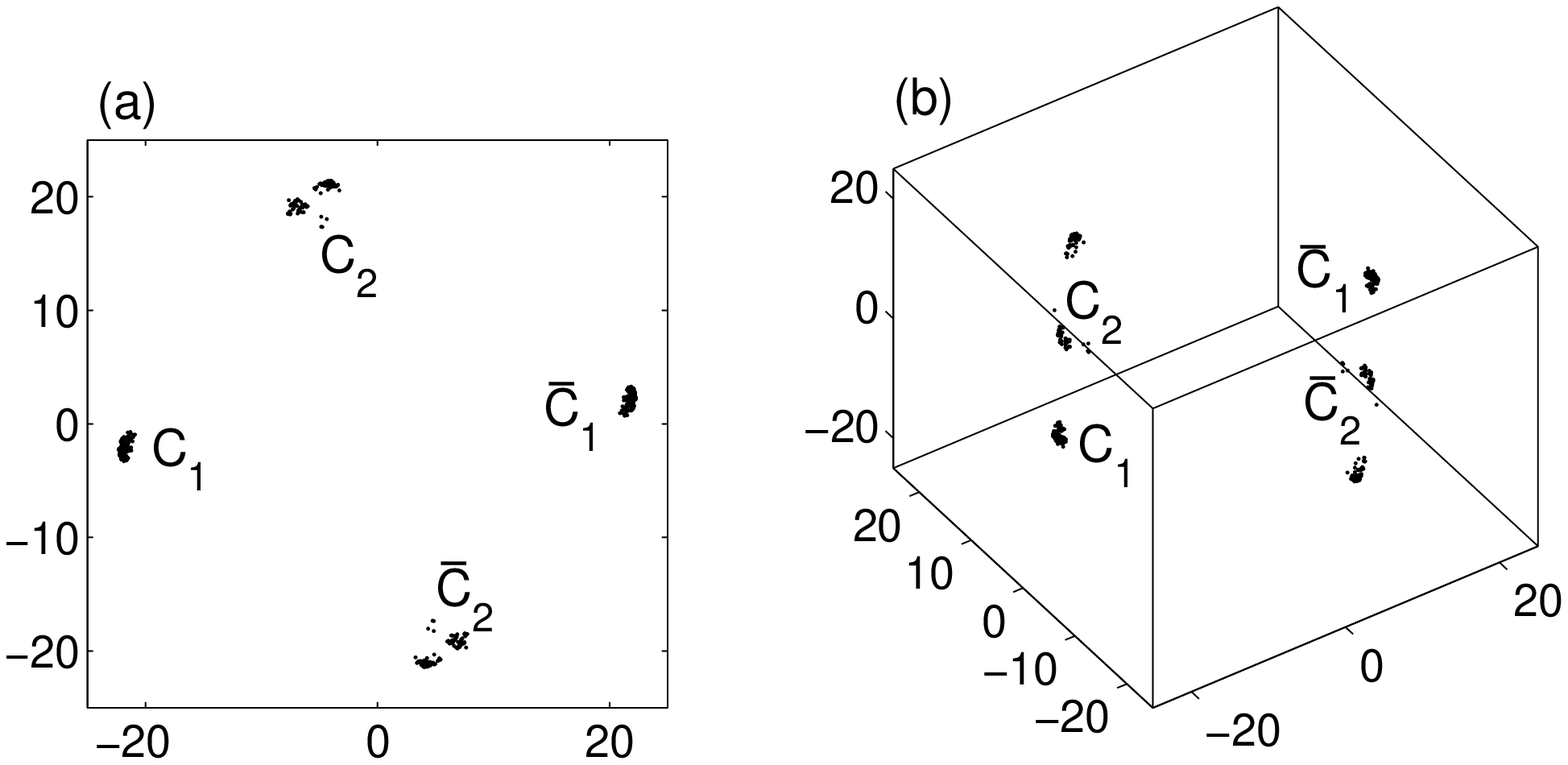,width=12cm}}
\vspace{2mm}
\caption{
Principal Component Analysis of a sample of $M=500\times2$ states
of a specific realization of $\{J\}$ in $3 D$
with $N=8^3$ spins at $T=0.2$. Each point represents a state
${\mathbf S}^\mu$. The coordinates are projections on to eigenvectors
{\bf e}$_i$, corresponding to the largest eigenvalues
of the correlation matrix in Eq.~(\ref{eq:Rstate}).
We show in (a) projections onto two eigenvectors, corresponding to the
largest and next-largest eigenvalues of the correlation matrix,
shown, respectively, on the
horizontal and  vertical axes.
In (b) the three largest eigenvectors are used.
The first
and second level partitions of the hierarchy
are clearly visible and, to some extent, the third level also.
}
\label{fig:pca}
\end{figure}

We describe here analysis of a single realization of the randomness,
in order to help the reader perceive the qualitative nature of the results
(see Figs. \ref{fig:datmat} and \ref{fig:pca}), and to define the
observables that we measure.
These observables were measured for each of the different realizations,
and the distributions of their values
were determined; the
average and width of these distributions are also presented. This data
demonstrate
that the results described in this section for a single sample are typical
and seen in  many samples.

In order to cluster the states,
each state $\mu$ is represented as
an $N$-component vector ${\mathbf S}^\mu = (S^\mu_1, ... ,S^\mu_N)$,
where $S_i^\mu = \pm 1$
is the value taken by spin $i$ in state $\mu$. The complete data set
can be represented as an $N \times M$ {\em data matrix}\/,
whose columns are the vectors
${\mathbf S}^\mu$. For the set of $M=1000$ states, obtained
at $T=0.2$ for a particular
bond realization of an $N=8^3$ spin system, the data matrix is
presented in Fig  \ref{fig:datmat}(a). Pixel $(i,\mu)$ of this figure
represents the sign of spin $i$ in state $\mu$; a black entry corresponds to
$+1$ and white to $-1$. The spins appear in lexicographic order and
the states in the random order generated by the simulation. As can be
seen, the matrix appears fairly random, with no easily discernible
structure; nevertheless, there is a clear organization of these $M$
states into
tight clusters. For the particular realization and ensemble of states
presented here, these clusters of states can be seen by direct observation
of the $M=1000$ data-points ${\mathbf S}^\mu$,
once one overcomes the hurdle of directly
viewing a cloud of 1000 points in a $N=512$ dimensional space.

A trivial
way of visualizing points that lie in a high dimensional space is to
project them onto a low (i.e. two or three) dimensional subspace.
In order to reveal the underlying structure, it is important to choose
with care the subspace onto which one projects.
A widely used method to choose this subspace is that of {\it principal
component analysis} (PCA)~\cite{PCA}.
One constructs the $N \times N$
covariance matrix of the $M$ points,
\begin{equation}
r_{ij}=\frac{1}{M}\sum_{\mu=1}^M \delta S^\mu_i \delta S^\mu_j
\end{equation}
where
\begin{equation}
\delta S^\mu_i = (S^\mu_i - m_i)/\sigma_i
\end{equation}
with $ m_i $ the average of the $M$ variables $S^\mu_i$ and $\sigma_i^2$ their
variance. For our case $ m_i = 0$ and $\sigma_i=1$ for all $i$, and hence the
covariance matrix is the spin correlation matrix, {\em i.e.}\/
\begin{equation}
r_{ij}=
c_{ij}=\frac{1}{M}\sum_{\mu=1}^M S^\mu_i S^\mu_j \;.
\label{eq:Rstate}
\end{equation}
The eigenvectors {\bf e}$_i$ of this matrix are
the principal directions or components of the variation in the data.
They are ordered according to the size of the corresponding
eigenvalues, with the largest coming first.

In  Fig. \ref{fig:pca}
we present the projections of our $T=0.2$ ensemble of $M=1000$
states on the   first two and three principal components.
Even though projection of $N=8^3$ dimensional data onto three and two
dimensions involves a major loss of information, the cluster structure
of the states is still clearly evident.
In Fig. \ref{fig:pca} (a) projection onto
the largest
eigenvector, {\bf e}$_1$, is represented by the horizontal axis,
and on the second largest, {\bf e}$_2$,
by the vertical.
It is interesting to
note that the two largest state clusters, $\C_1$ and ${\bar \C_1}$,
project mostly onto {\bf e}$_1$ and the
second largest pair, $\C_2$ and ${\bar \C_2}$
onto {\bf e}$_2$. Fig.  \ref{fig:pca} (b)
indicates that the next sized variation, due
to splitting of
$\C_2$ into two subgroups, is captured by {\bf e}$_3$. The scale of the
projections can be understood by the following argument:
if the (normalized)
eigenvector {\bf e}$_1$ is
parallel to a typical vector from $\C_1$, then, since
normalization
of {\bf e}$_1$ involves a factor of $1/\sqrt{N}$, the maximum
possible projection is $\sqrt{N} \approx 22.6$. Hence the projections shown in
Fig. \ref{fig:pca} are quite large, i.e. close to the maximum possible value.

Next we obtain a systematic quantitative measure of the hierarchical structure
of state space by performing a cluster analysis of the $M$ points.
The choice of the particular clustering algorithm used was dictated by
our idea of the state space structure, obtained from PCA and from the
picture described in the Introduction and
summarized in Fig. \ref{fig:scheme}.
Our aim is to find a hierarchy of partitions into compact clusters.
That is, we would like states that belong to the same cluster to be closer
to each other than to states in different clusters.
Ward's algorithm, described in Section \ref{sec:clustering}, is tailored
to perform  this task for the kind of data distribution that we have
in state space.

To start, we
defined the $M\times M$ distance matrix 
$D$ between the states $\mu, \nu$
by
\begin{equation}
D_{\mu\nu} = {1 - q_{\mu\nu} \over 2} \;,
\label{eq:dmn}
\end{equation}
where $q_{\mu\nu}$ is the state overlap defined by Eq.~(\ref{eq:q}).
Next, we clustered the states using the distance matrix $D_{\mu\nu}$ as input
to Ward's algorithm (see Eq.~(\ref{eq:Ward})).
The algorithm results in a dendrogram, as shown in Figs.
\ref{fig:state_dend}(a,c,e), for a sample at $T=0.2,0.5$ and $2.0$, in
three dimensions.
The leaves, which represent the states, are ordered on the
horizontal  axis according to the order imposed by the dendrogram
\cite{Alon99}. The nodes represents the clusters.
The vertical location of each cluster corresponds to its $\tau$ value,
and is thus related to the variance within it.

For $T=0.2$ and $0.5$, which are below $T_c \approx 0.95$ \cite{Marinari98},
we found clear partitions in the two highest levels
of the dendrogram, as presented in Figs. \ref{fig:state_dend}(a,c).
At the highest level the states are partitioned into
$\C$ and $\bar\C$. At the next level, $\C$ is broken into two sub-clusters,
which we denote as $\C_1$ and $\C_2$. For this specific
sample the
cluster $\C_2$ breaks further into two sub-clusters, which are clearly seen in
Fig. \ref{fig:pca} as well.

To gain insight into the manner in which similar states are grouped together,
and to actually ``look into the spin-glass'' at the microscopic level,
we present in Fig. \ref{fig:datmat}(b) the same data matrix as
shown in Fig. \ref{fig:datmat}(a), but with the states again reordered according to
the dendrogram of Fig. \ref{fig:state_dend} (a).
That is, to get Fig. \ref{fig:datmat} (b),
the columns of  Fig. \ref{fig:datmat}(a) have been
permuted according to their position in the dendrogram.
The clear central vertical dividing line separates $\C$
from $\bar \C$.
In addition to the central dividing line, another vertical line is also
clearly visible - it separates the states that belong to the larger cluster
$\C_1$ from the smaller one, $\C_2$.

We now demonstrate that 
the
state clusters we found are indeed ``correct'' and ``natural''.
First, we checked that the situation of merging two clusters of very different
sizes occurs very rarely.

We showed that our partitions are ``natural'' 
and not an artefact of the algorithm
(which produces a tree for any data), 
in three ways:
\begin{enumerate}
\item
Note that direct observation of the dendrograms clearly distinguishes
between the different situations
above and below $T_c$. At $T=0.2,0.5 \ (< T_c)$ the relative $\tau$ values
of the state clusters $\C,\C_1$ and $\C_2$ -- measured by the length of the
branch above
each cluster -- are high. A long branch indicates that the size of
the cluster is much smaller than the distance between it and its ``brother",
which indicates that the partition into these two groups is natural. In
comparison, in the dendrogram obtained at $T=2.0\ (> T_c)$,
the relative $\tau$ values are much smaller than at $T=0.2,0.5$.

\item
The
genuinely hierarchical structure at $T=0.2,0.5$ is also evident
from the states' distance matrix, as shown in Figs. \ref{fig:state_dend}(b,d).
This distance matrix was obtained by reordering the states according to the
results of the cluster analysis, i.e. according to the order of the leaves of
the corresponding dendrogram. When the states are randomly ordered
(like in Fig. \ref{fig:datmat}(a)),
the resulting distance matrix is a homogeneous greyish
square, like that of Fig. \ref{fig:state_dend}(f). The difference between this
and Figs. \ref{fig:state_dend}(b,d) is striking: the distance matrices {\it
within} clusters $\C_1$ and $\C_2$ appear as dark
squares (representing shorter distances)
along the diagonal.
The distances
between clusters are represented by fairly uniform, lighter colored rectangles.
In comparison, for $T=2.0$ there is no real hierarchical organization of
the states, and reordering them according to the dendrogram does not generate
any ordered appearance of the  distance matrix.

\item
We
measured the average distance between pairs of states that
belong to each of the
clusters $\C$, $\C_1$ and $\C_2$. The average $D(\C)$ and the width
$w(\C)$ of the distribution of distances within $\C$ are
\begin{eqnarray}
&D(\C) = {1\over|\C|^2} \sum_{\mu,\nu\in\C} D_{\mu\nu} \;;
\label{eq:dbc}\\
&w(\C) = \left({1\over|\C|^2}\sum_{\mu,\nu\in\C}{D_{\mu\nu}}^2 - D(\C)^2
\right)^{1/2} \;,
\label{eq:wbc}
\end{eqnarray}
where $\mu$ and $\nu$ refer to individual configurations.
The average $D(\C_\alpha)$ and the width $w(\C_\alpha)$ for $\alpha=1,2$
are defined in a similar way.
The distribution of distances within clusters is to be compared with
the distribution of distances
between points that belong to different clusters. The average $D(\C_1,\C_2)$
and width $w(\C_1,\C_2)$
of the inter-cluster distance distribution are defined as
\begin{eqnarray}
&D(\C_1,\C_2) =
{1\over|\C_1||\C_2|} \sum_{\mu\in\C_1}\sum_{\nu\in\C_2} D_{\mu\nu} \;;
\label{eq:dcc} \\
&w(\C_1,\C_2) = \left( {1\over|\C_1||\C_2|} \sum_{\mu\in\C_1}\sum_{\nu\in\C_2}
{D_{\mu\nu}}^2 -D(\C_1,\C_2)^2 \right)^{1/2} \;.
\label{eq:wcc}
\end{eqnarray}
The clusters $\C, \bar \C$ are special in that
each state $\mu\in\C$ has an inverted state ${\bar \mu}\in\bar\C$, so that
${\mathbf S}^\mu=-{\mathbf S}^{\bar \mu }$. Therefore $D(\C,\bar\C)=1-D(\C)$
and $w(\C,\bar\C)=w(\C)$.

A subset of the
results is presented in Table \ref{tab:dcc}; for all  temperatures, system 
sizes and both dimensions see~\cite{HedThesis}. 
We present  for each variable $x$ its mean $[x]_J$ (averaged
over the disorder $\{J\}$)
and its standard deviation $\Delta x = ([x^2]_J-{[x]_J}^2)^{1/2}$.
For $T = 0.2$ and $0.5$, which are below $T_c$,
the average distances within the clusters are of the order
of 0.1. $D(\C,\bar\C)$ is around $0.9$, which shows that there is a clear
separation
between these two clusters. $D(\C_1,\C_2)$ is much lower, but is still
about two or three times larger than either $D(\C_1)$ or $D(\C_2)$. Note that
the width of the distance distribution within a cluster is of the same order
of the mean distance, so in general distances will not be much larger than
twice the mean distance.
At $T=2.0\ (>T_c)$
the distances within and between clusters are almost equal and
the differences are only due to statistical fluctuations, again indicating
absence of natural structure, as we claimed on the basis of direct observation.

\end{enumerate}


\begin{figure}
\centerline{\psfig{figure=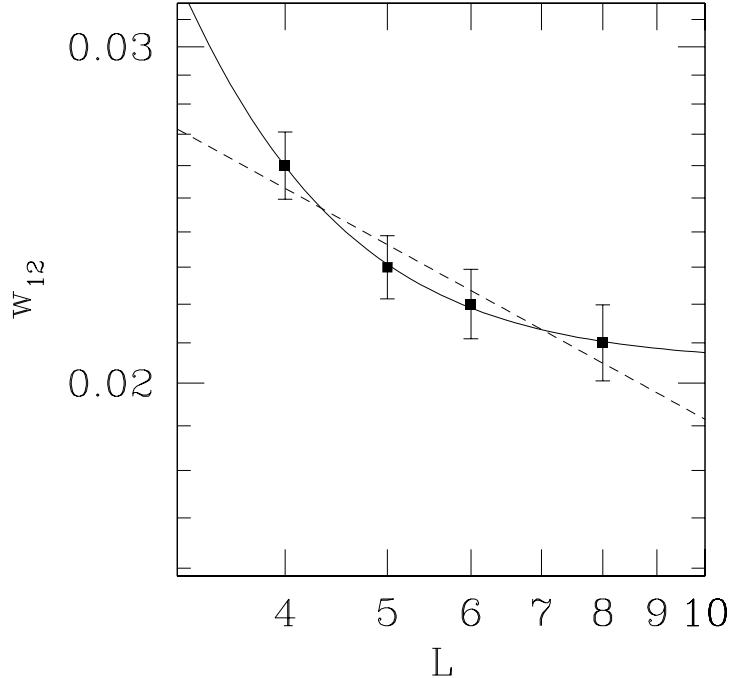,width=10cm}}
\vspace{2mm}
\caption{
A log-log plot of $w_{12}$ against $L$ for $T=0.2$ and $D=3$. The solid line
is the best least squares fit to Eq.~(\ref{eq:w12fit}),
while the dashed line is the
best fit with the additional assumption that $w_\infty = 0$. .
}
\label{fig:w12}
\end{figure}

Measurement of some of the quantities listed above allows us to
investigate the extent to which the state space structure of short-range spin
glasses, as reflected by 
the data in Table \ref{tab:dcc}, is
compatible with RSB.
In the RSB \cite{Parisi79,Parisi80,Parisi83,RSB}
framework, the overlap between any pair of valleys (which correspond
to pure states in the usual interpretation of RSB) from two different 
clusters that appear at the 
same level of the hierarchy is constant.
It seems natural to associate the pure state clusters of RSB to our
state clusters, e.g. $\C_1$ and $\C_2$. In this association,
each state cluster
contains states that belong to different ``pure states''.
If the overlap between pure states of the two clusters is constant 
as in RSB, this should hold
also for the overlap between each pair of states $\mu\in\C_1$ and $\nu\in\C_2$,
since the width of the overlap distribution inside a pure state 
approaches zero.
In this case, all entries of
the sub-matrix $\tilde D_{\mu \nu}$ for $\mu \in\C_1$ and $\nu \in\C_2$
would be equal,
so the width $w_{12}=[w(\C_1,\C_2)]_J$
should vanish as $L\rightarrow\infty$. To test whether this is the
case, we present in Fig. \ref{fig:w12} the values of
$w_{12}=[w(\C_1,\C_2)]_J$ {\it vs} the system size $L$ for $T=0.2$
and $D=3$. The error bars represent the statistical error
(obtained by dividing the standard deviations, given in Table
\ref{tab:dcc}, by $\sqrt{N_{\rm samp}-1}$).
We tried fits of the form
\begin{equation}
w_{12} = w_\infty + B L^{-y} \;,
\label{eq:w12fit}
\end{equation}
with $B$ and $y$ as
fit parameters. The overall best fit was for $ w_\infty = 0.0205, B = 0.58$ and
$y = 3.36$, which gives a very small $\chi^2$ of $0.036$. This is shown by
the solid line in  Fig. \ref{fig:w12}. We also tried the best
fit assuming that
$w_\infty = 0$, which has fit parameters $B = 0.039$ and
$y =  0.30$, and is shown by the dashed line in the figure. This has a
$\chi^2$ of $1.41$ which is much larger than the best fit with $w_\infty \ne 0$,
but still acceptable. Hence even though our data suggests 
that $w_\infty \ne 0$, the
possibility that $w_\infty = 0$, which corresponds to RSB, cannot be ruled
out.

\begin{table}[p]
\begin{tabular}[t]{|ll|llll|llll|}
\hline
 $T$ & $L$ & $[D(\C)]_J$     & $\Delta D(\C)$   & $[w(\C)]_J$   & $\Delta w(\C)$
                & $[D(\C_1)]_J$   & $\Delta D(\C_1)$ & $[w(\C_1)]_J$ & $\Delta w(\C_1)$    \\ \hline
0.2& 4  & 0.045 & 0.049 & 0.055 & 0.052 & 0.015 & 0.017 & 0.019 & 0.018\\
   & 5  & 0.050 & 0.054 & 0.056 & 0.054 & 0.018 & 0.018 & 0.019 & 0.019 \\
   & 6  & 0.053 & 0.056 & 0.054 & 0.053 & 0.021 & 0.020 & 0.019 & 0.019 \\
   & 8  & 0.055 & 0.054 & 0.052 & 0.051 & 0.025 & 0.020 & 0.020 & 0.020 \\ \hline
0.5& 8  & 0.139 & 0.065 & 0.084 & 0.046 & 0.093 & 0.038 & 0.045 & 0.026 \\
   & 12 & 0.151 & 0.065 & 0.078 & 0.046 & 0.106 & 0.036 & 0.041 & 0.024 \\ \hline
2.0& 8  & 0.487 & 0.006 & 0.053 & 0.002 & 0.477 & 0.009 & 0.055 & 0.002 \\ 
\hline \hline
   &    & $[D(\C_2)]_J$       & $\Delta  D(\C_2)$     & $[w(\C_2)]_J$      & $\Delta w(\C_2)$
                & $[D(\C_1,\C_2)]_J$  & $\Delta D(\C_1,\C_2)$ & $[w(\C_1,\C_2)]_J$ & $\Delta w(\C_1,\C_2)$ \\ \hline
0.2& 4  & 0.025 & 0.036 & 0.027 & 0.034 & 0.160 & 0.135 & 0.026 & 0.024 \\
   & 5  & 0.025 & 0.032 & 0.025 & 0.031 & 0.169 & 0.147 & 0.023 & 0.020 \\
   & 6  & 0.028 & 0.033 & 0.026 & 0.033 & 0.161 & 0.141 & 0.022 & 0.021 \\
   & 8  & 0.030 & 0.027 & 0.024 & 0.026 & 0.161 & 0.139 & 0.021 & 0.018 \\ \hline
0.5& 8  & 0.112 & 0.057 & 0.057 & 0.037 & 0.253 & 0.126 & 0.053 & 0.027 \\
   & 12 & 0.121 & 0.048 & 0.054 & 0.033 & 0.263 & 0.125 & 0.044 & 0.023 \\ \hline
2.0& 8  & 0.472 & 0.009 & 0.057 & 0.002 & 0.499 & 0.005 & 0.048 & 0.003 \\ 
\hline
\end{tabular}
\caption{The average distances within and between state clusters,
and the relations between them, for a subset of the $D=3$ dimensional systems. 
For each variable $x$ we present the average over all realizations, $[x]_J$,  followed by its
standard deviation, i.e. $\Delta x  =  ([x^2]_J - {[x]_J}^2)^{1/2}$.
The statistical error of each mean $[x]_J$ is $\Delta x / \sqrt{N_{\rm samp}}$; the number of
samples for each $L,D$ is given in Table \ref{table:34d}.
}
\label{tab:dcc}
\end{table}

\section{Correlated domains in spin space}
\label{sec:spins}

\ssection{Identifying the spin domains}
\label{sec:spindomains}

According to our picture, splitting of a cluster at level $a$ in
the states hierarchy is induced by a macroscopic contiguous~\cite{foot7}
 spin
domain $\G_a$. 
The size and shape of this domain determines the
energy barrier
separating two state
clusters that were ``born'' at this level.  In this subsection we
describe how we identify from our data the two correlated domains
$\G_1$ and $\G_2$, which determine the two highest levels of the
states hierarchy, and also discuss whether they remain macroscopic
at large $L$.
Domains that emerge at the
next level, $\G_3$ and $\G_3^\prime$, are also discussed briefly.

Since the spins in such  domains flip ``collectively'', they are
highly correlated. The standard definition of the correlation
$c_{ij}$ of spins $i$ and $j$ is \beq c_{ij}= \langle S_i S_j
\rangle = \frac{1}{\cal Z} \sum_{\mathbf S} S_i S_j
\exp[-\beta{\cal H}({\mathbf S})] \;, \label{eq:cij} \eeq where
$\langle ... \rangle$ stands for the thermodynamic average for a
particular realization of the disorder, and ${\cal Z}$ is the
partition function at $T$. Using our equilibrium ensemble of
states $\{{\mathbf S}^\mu\}$, we evaluate \beq c_{ij}={1\over
M}\sum_\mu S_i^\mu S_j^\mu \;. \eeq The correlation in itself is
unimportant for spin glasses since it is gauge dependent and its
average $[c_{ij}]_J$ over all the realizations of the disorder
$\{J\}$ vanishes. The relevant measure of correlations in a spin
glass is the square, ${c_{ij}}^2$. If two spins are independent of
each other over the equilibrium ensemble of states, we have
${c_{ij}}^2=0$. On the other hand, for a pair of fully correlated
spins we have ${c_{ij}}^2=1$; the two spins are either aligned or
anti-aligned in all states.


To proceed,
it is convenient to define, quite generally, $\G_{\mu \nu}$ as the set of spins
whose sign is different in $\mu$ and $\nu$, {\em i.e.}\/
\begin{equation}
\G_{\mu\nu} = \left\{\; i \;|\; S_i^\mu \not= S_i^\nu \; \right\} \;.
\label{eq:gmn}
\end{equation}
We expect the largest domain, $\G_1$, to be in one orientation in
the states of $\C$ and in the reversed one in the states of $\bar
\C$. To identify the spins that indeed behave this way, we took
all $(M/2)^2$ pairs of states  $\mu\in\C$ and $\nu\in\bar\C$ and,
for each pair, determined $\G_{\mu\nu}$. Ideally all the spins of
$\G_1$  always flip together and maintain their relative
orientation; if so, the set of spins $\G_{\mu\nu}$ for all pairs of
states $\mu$ and $\nu$ would always include $\G_1$.
However, at finite $T$ we must allow for excitations of
the order of $J$. So, even if a spin is highly correlated with the
other spins of $\G_1$, it might lose its relative orientation in a
few of the $M$ states of the sample. In order not to ``miss'' such
spins, we use a soft criterion when we determine whether a spin is
a member of $\G_1$. We define a threshold $\theta$ and define
$\tilde\G_1(\theta)$ as the set of spins $i$ which are members of
$\G_{\mu\nu}$, i.e. for which $S^\mu_iS^\nu_i=-1$, for at least a
fraction $\theta$ of the pairs of states $\mu\in\C$ and
$\nu\in\bar\C$. This can be written as
\beq \label{eq:tg}
\tilde\G_1(\theta) = \left\{ \; i \; \left| \;
{1\over|\C||\bar\C|} \sum_{\mu\in\C} \sum_{\nu\in\bar\C} S_i^{\mu}
S_i^{\nu}  < 1-2\theta \right. \right\} \;,
\eeq
since the terms
in the normalized sum where $S_i^{\mu} S_i^{\nu}=1$ must, by
definition, sum up to less than $1 - \theta$ and the sum of the terms with
$S_i^{\mu} S_i^{\nu}=-1$ must be less than $-\theta$.
We define
our spin domain $\G_1(\theta)$ as the largest contiguous part of
$\tilde\G_1(\theta)$. For large enough $\theta$ we found that for
most realizations $\{ J \}$, below $T_c$  the sites of $\tilde\G_1(\theta)$ are
contiguous and hence it is identical to $\G_1(\theta)$
(for detailed values of the ratio $|\G_a |/ |\tilde\G_1 | $, its mean over
realizations and its standard deviation, see ~\cite{HedThesis}). 
The next spin domain $\G_2(\theta)$ is
defined in the same manner, on the basis of pairs of states $\mu
\in \C_1$ and $\nu \in \C_2$.

The above definition sets a lower bound on the correlation of spins within the
domain.
Consider two spins $i,j\in\G_1(\theta)$. By definition,
\begin{equation}
{c_{ij}}^2
={1\over M^2} \sum_{\mu,\nu} S^\mu_i S^\mu_j S^\nu_i S^\nu_j .
\end{equation}
Now the number of states in $\C$ and $\bar\C$ are both equal to $M/2$. In
addition, for a given $\nu$, we can replace $\mu$ by its
inverse $\bar\mu$ and the product of the four spins doesn't change.
Hence we get the same contribution from $\mu\in\C$ as $\bar\mu\in\bar\C$.
As result we have
\begin{equation}
{c_{ij}}^2
 ={1\over |\C||\bar\C|} \sum_{\mu\in\C} \sum_{\nu\in\bar\C}
S^\mu_i S^\mu_j S^\nu_i S^\nu_j  \;.
\label{eq:c_C_Cbar}
\end{equation}
Now $S^\mu_i S^\nu_i $ will be $-1$ for a fraction of the states $\mu$ and
$\nu$ which is greater than $\theta$ and $+1$ for a fraction less than
$1-\theta$, and similarly for $S^\mu_j S^\nu_j $. Hence $S^\mu_i S^\nu_i $ and
$S^\mu_j S^\nu_j $ will have the
same sign with probability greater than $1-2(1-\theta) = 2\theta-1$.
Consequently, for $i,j\in\G_1(\theta)$, we have
\begin{equation}
{c_{ij}}^2  >  2\theta-1 - \left[1 - (2\theta-1)\right] = 4\theta -3 .
\label{eq:gcorr}
\end{equation}
The same constraint holds also for $\G_2$, with the sums taken over the
states in clusters $\C_1$ and $\C_2$.

\begin{figure}[t]
\centerline{\psfig{figure=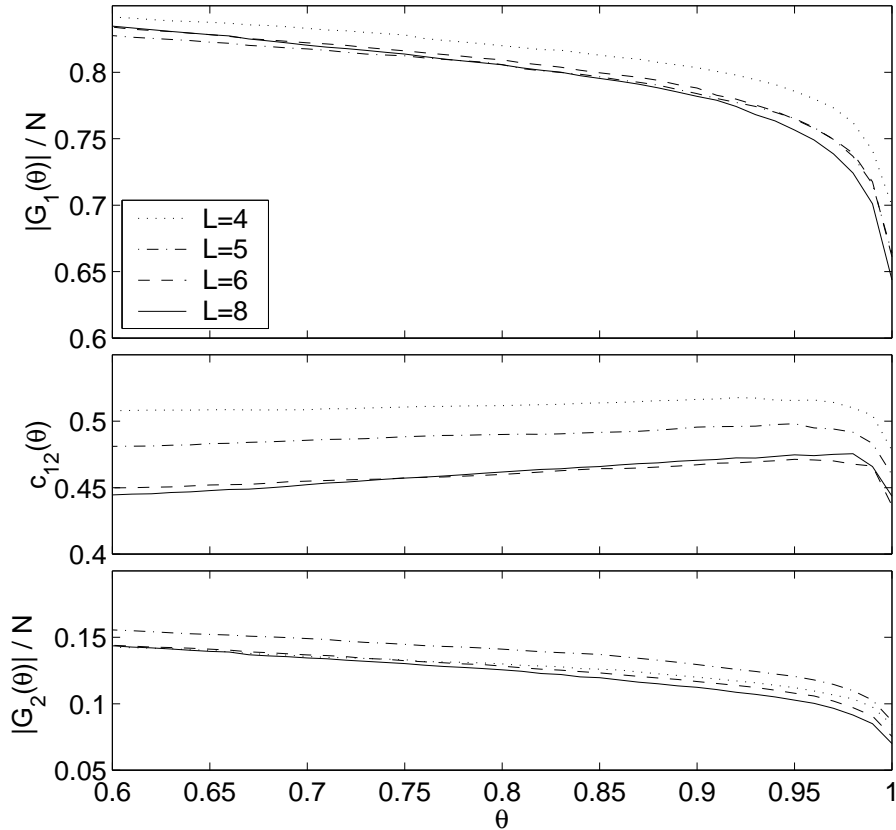,width=12cm}}\vspace{2mm}
\caption{The normalized sizes of the two largest spin domains,
$\G_1(\theta)/N$ and $\G_2(\theta)/N$ and their
correlation $\bar c_{12}$, defined in Eq.~(\ref{eq:c12}),
as a function of the threshold $\theta$
for $D=3$, $T=0.2$.
}
\label{fig:gvsth}
\end{figure}

\begin{figure}[t]
\centerline{\psfig{figure=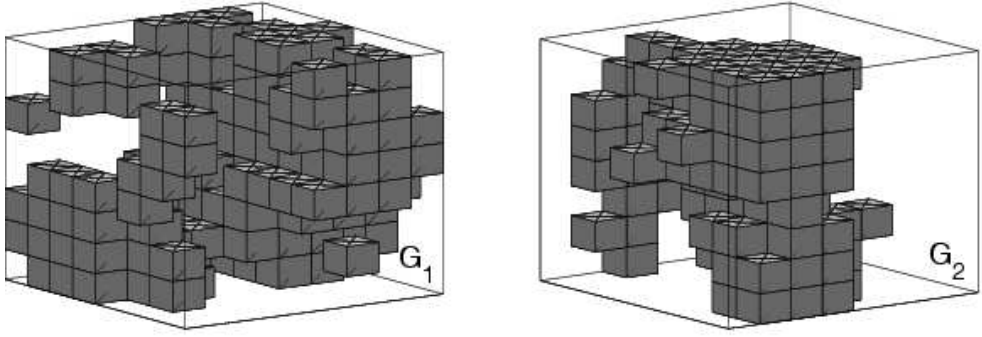,width=13cm}}
\vspace{2mm}
\caption{The spin domains $\G_1$ and $\G_2$, as found in the realization
of Fig. \ref{fig:datmat}. Note that we use periodic boundary conditions, so
the domains are connected through the boundaries. No spin is shared by $\G_1$
and $\G_2$.}
\label{fig:sponges}
\end{figure}

Since we introduced an arbitrary  parameter $\theta$ into the definition of
our spin clusters, it is important to consider the extent to
which the value of $\theta$ affects their identification.
As seen in Fig. \ref{fig:gvsth}, the sizes of the domains and their average
correlation, defined below in (\ref{eq:c12}), do not change
much for $0.6 \leq \theta \leq 0.95$. For both $a=1,2$ we define
(arbitrarily) $\G_a=\G_a(0.95)$. We do not choose $\theta=1$ since,
as discussed above,
we do not want our results to be affected by small thermal fluctuations.
In Fig. \ref{fig:sponges} we plot the spatial structure of $\G_1$ and $\G_2$ for
a specific realization.
For $T>T_c$ the correlations between each pair of spins are much smaller,
and hence this analysis
is meaningless. The procedure described above results in
$\G_1(\theta)=\G_2(\theta)=\emptyset$ for any $\theta>0.5$.

\begin{figure}[t]
\centerline{\psfig{figure=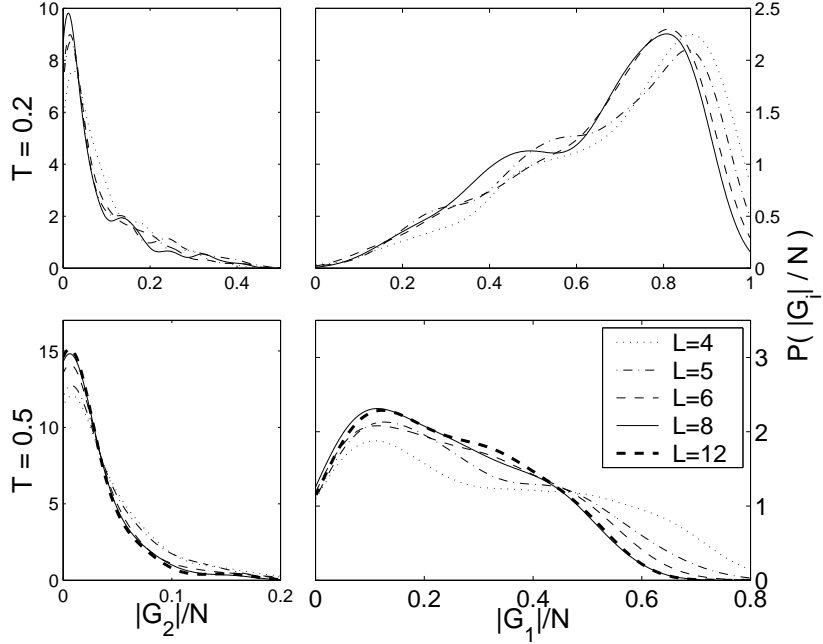,width=11cm}}
\caption{Size distributions of the spin domains $\G_1$ and $\G_2$
for $D=3$ dimensions at $T=0.2,0.5$. The distributions seem to
converge, despite the small system sizes.} \label{fig:gg3d}
\end{figure}

\begin{figure}[t]
\centerline{\psfig{figure=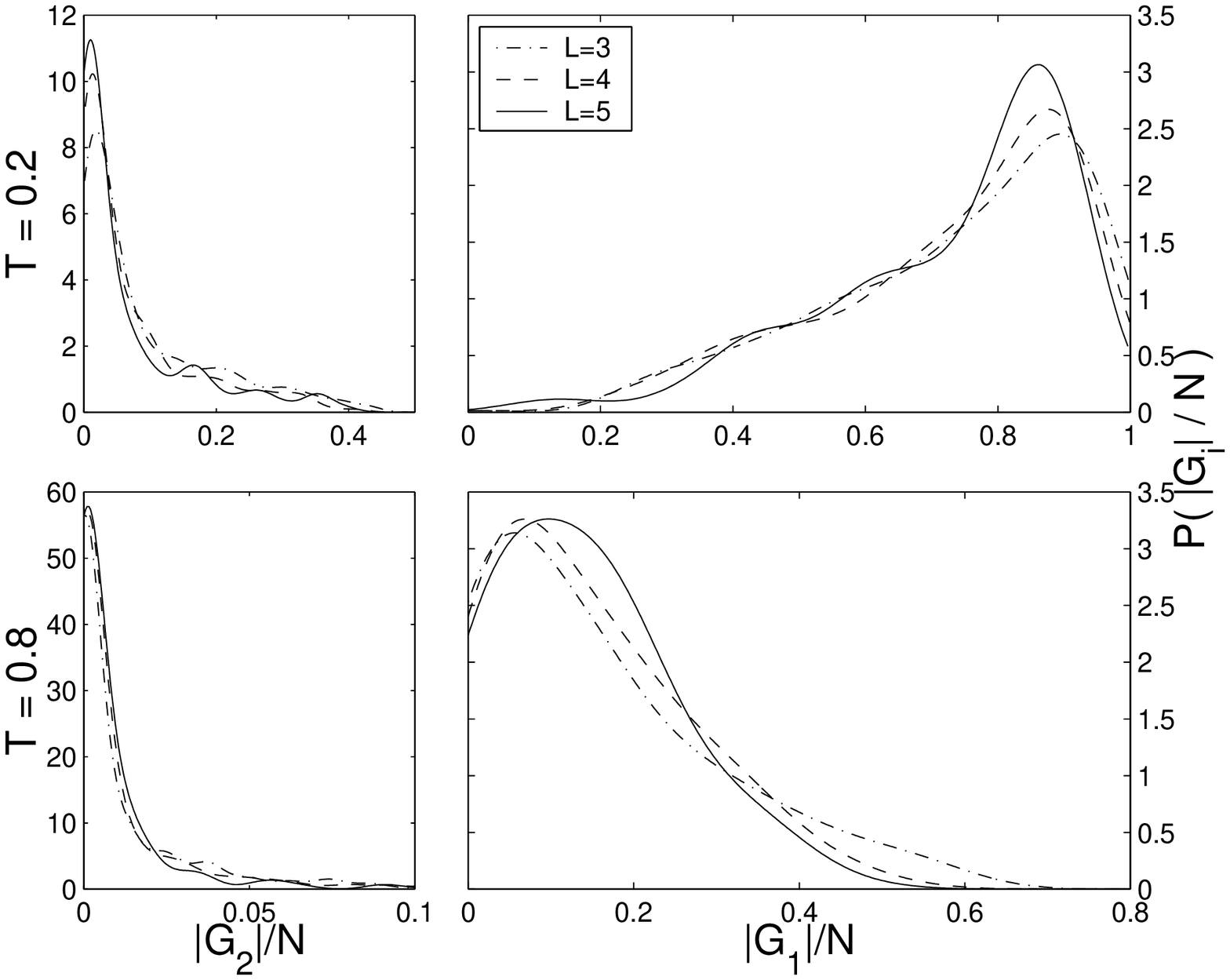,width=11cm}}
\caption{Size distributions of the spin domains $\G_1$ and $\G_2$ for $D=4$
dimensions at $T=0.2,0.8$. For $T=0.2$ the distributions seem to
converge, despite the small system sizes. For $T=0.8$, $|\G_2|/N$ converges to
a narrow distribution around zero, and $|\G_1|/N$ does not show convergence
yet.}
\label{fig:gg4d}
\end{figure}

According to our picture these correlated spin domains govern the
hierarchical structure of state space. It is important to clarify
whether these domains survive as the system size $L$ increases.
There are two mechanisms by which increasing the system size can
invalidate our picture: either the domains do not remain macroscopic
when $L$ increases, or they
do remain macroscopic
but merge as $L\rightarrow\infty$, i.e. the fraction of states in which $\G_2$
flips tends to zero. We now discuss each of these possibilities
in turn. In addition,
a simple figurative
description of these two mechanisms is given Sec.~\ref{sec:spincluster}.

\begin{enumerate}
\item
The domains do not remain macroscopic
when $L$ increases. To study the finite
size effects of our analysis we normalized the domain sizes by the
number of spins and plotted the size distributions of the two
domains for different system sizes, in $D=3$ (see Fig.
\ref{fig:gg3d}) and in $D=4$ (Fig. \ref{fig:gg4d}), at two
temperatures in both dimensions. The number of
bond realizations, $N_{\rm samp}$,
from which these distributions were obtained for
various system sizes $L$, at both $D=3,4$, are given in Table
\ref{table:34d}.
For $T=0.2$ in both dimensions,
and at $T=0.5$ for $D=3$ the distributions seem to converge even
for the small system sizes we use. We conclude with high certainty
that at $T=0.2$ for $D=3,4$ and at $T=0.5$ for $D=3$ the domain
sizes $\vert \G_a \vert $ are proportional to $ L^D$ for both $a=1,2$.
The mean
and width of these distribution are presented in Table \ref{tab:c12}.
The width of the distributions does not vanish,
so the sizes of the domains are non self-averaging quantities.
On the other hand, for
$T=0.8$ in $D=4$ we cannot determine conclusively whether the
domain sizes do or do not remain proportional to $N=L^D$ as $L$
increases.

\begin{table}[p]
\begin{center}
\begin{tabular}[t]{|lll|lllllll|}
$D$ & $T$ & $L$ & $[|\G_1|]_J/N$ & $\Delta|\G_1|/N$ & $[|\G_2|]_J/N$ & $\Delta|\G_2|/N$ & $[\bar
c_{12}]_J$ & $\Delta\bar c_{12}$ & $P(\G_2\not=\emptyset)$ \\ \hline
{\bf 3}&0.2& 4  & 0.70(1) & 0.21 & 0.099(4) & 0.087 & 0.56(1) & 0.33 & 0.856(6) \\
    &     & 5   & 0.66(1) & 0.21 & 0.105(5) & 0.104 & 0.55(1) & 0.33 & 0.832(6) \\
    &     & 6   & 0.66(1) & 0.20 & 0.090(4) & 0.090 & 0.52(2) & 0.34 & 0.836(6) \\
    &     & 8   & 0.64(1) & 0.20 & 0.084(5) & 0.094 & 0.53(2) & 0.34 & 0.833(8) \\
\hline
    & 0.5 & 4   & 0.31(1) & 0.21 & 0.062(3) & 0.056 & 0.49(1) & 0.32 & 0.56(1) \\
    &     & 5   & 0.26(1) & 0.18 & 0.052(2) & 0.043 & 0.49(1) & 0.33 & 0.57(1) \\
    &     & 6   & 0.25(1) & 0.16 & 0.046(2) & 0.046 & 0.47(1) & 0.33 & 0.52(1) \\
    &     & 8   & 0.22(1) & 0.15 & 0.035(2) & 0.034 & 0.47(2) & 0.31 & 0.55(1) \\
    &     & 12  & 0.24(1) & 0.15 & 0.033(2) & 0.035 & 0.54(2) & 0.31 & 0.56(2) \\
\hline
{\bf 4}&0.2& 3  & 0.74(1) & 0.19 & 0.107(5) & 0.105 & 0.62(2) & 0.34 & 0.840(6) \\
    &     & 4   & 0.73(1) & 0.19 & 0.083(4) & 0.092 & 0.53(2) & 0.34 & 0.830(6) \\
    &     & 5   & 0.73(1) & 0.19 & 0.082(7) & 0.098 & 0.51(2) & 0.34 & 0.77(1) \\
\hline
    & 0.8 & 3   & 0.154(7) & 0.15 & 0.036(1) & 0.031 & 0.47(1) & 0.31 & 0.298(9) \\
    &     & 4   & 0.142(6) & 0.12 & 0.025(1) & 0.029 & 0.54(1) & 0.31 & 0.37(1) \\
    &     & 5   & 0.139(8) & 0.11 & 0.020(2) & 0.025 & 0.57(2) & 0.29 & 0.38(2) \\
\hline
\end{tabular}
\end{center}
\caption{
The normalized sizes of
the domains $\G_1$ and $\G_2$, and the average correlation between
spins that belong to the two domains. The last two parameters are
taken for realizations $\{J\}$ where $\G_2$ does not vanish. The
probability for $\G_2$ not to vanish is also presented. For each
quantity $x$ the table contains $[x]_J$, its average  over $N_{\rm samp}$
realizations of the disorder $\{J\}$ and the width of the distribution
$\Delta x = \sqrt{ [x^2]_J -{[x]_J}^2 }$. Next to each $[x]_J$ we
show its statistical error (in parentheses).}
\label{tab:c12}
\end{table}

\item
$\G_1$ and $\G_2$ may remain macroscopic but merge
as $L\rightarrow\infty$. If this occurs, we end up
with a single domain and there will be no hierarchical structure in
state space. To check that this does not happen we calculated the average
correlation $\bar c_{12}$ between spins in $\G_1$ and $\G_2$,
\beq
\bar c_{1 2}= {1\over|\G_1||\G_2|} \sum_{i\in\G_1}\sum_{j\in\G_2}{c_{ij}}^2 \;.
\label{eq:c12}
\eeq
If $\bar c_{1 2}$ approaches the value 1 as $L \rightarrow\infty$,
the two domains indeed merge in the thermodynamic limit.
In Table \ref{tab:c12} we present, for systems of different
sizes and dimensions, the average values of $\bar c_{12}$ (averaged over
the disorder $\{J\}$) and the corresponding
standard deviations. For $T=0.2,~D=3,4$ and for
$T=0.5,~D=3$ the average correlation  decreases slightly as the system size
increases,
although, in $D=3$ it seems to converge already for $L=8$
to a fixed value of $\sim0.5$.
This means that the spins of $\G_1$ and $\G_2$ will not become fully
correlated and the two domains will stay separate as $L$ increases.

Interestingly, in $D=4$,
the correlation for $L=4,5$ is higher at $T=0.8$ than at $T=0.2$.
The reason for this is probably that as $T$ increases, small
pieces  of $\G_1$ ``fall of''.
Since $\G_2$ at $T=0.2$ is small, one of these
pieces, which is larger than $\G_2$, assume the role of $\G_2$ at $T=0.8$.
Since this piece was part of $\G_1$ at $T=0.2$, we expect its correlation,
with what remains of $\G_1$ at $T=0.8$, to be relatively high.
Extrapolating from $L=3,4,5$ is not useful, but we still believe that the
correlation does not approach 1 as $L\rightarrow\infty$.
\end{enumerate}

We also attempted to identify $\G_3$ and $\G_3'$, the spin
domains associated with the third level of the state hierarchy
(see below). $\G_3$ is the cluster which is associated with
splitting $\C_1$ into its two descendents on the dendrogram,
$\C_{1a}$ and $\C_{1b}$. The domain $\G_3'$ plays the same role in
$\C_2$.
Since by our notation $|\C_1| \geq
|\C_2|$ we expected that in order to have a larger number of
states, the spin correlations  will be lower when measured over
$\C_1$ than over $\C_2$.
As a
result we expect $|\G_3| \leq |\G_3'|$. Due to the small sizes of
the systems we study, we cannot be sure if the sets of spins we
identify as $\G_3$ and $\G_3'$ indeed play the role we attribute
to them, or are just a microscopic noise and, therefore, only a
finite size effect. The results are given in Table \ref{tab:g3}.
We see that the normalized sizes of both domains decrease with the
system size, perhaps due to finite size effects. We also measure
the average correlation $\bar c(\G_3,\G_1\cup\G_2)$, of $\G_3$
with the largest domain correlated over $\C_1$, which includes
$\G_1\cup\G_2$ (this domain has a fixed orientation over the
states of $\C_1$). This correlation is defined as \beq \bar
c(\G_3,\G_1\cup\G_2)= {1\over|\G_1\cup\G_2||\G_3|}
\sum_{i\in\G_1\cup\G_2}\sum_{j\in\G_3}{c_{ij}}^2 \;.
\label{eq:c123} \eeq In Table \ref{tab:g3} we see that the values
of $\bar c(\G_3,\G_1\cup\G_2)$ decrease as $L$ increases; hence if
$\G_3$ survives as a macroscopic cluster at large $L$, we expect
it to remain distinct from the union of the two larger domains.

\begin{table}[p]
\begin{tabular}{|cc|ccc|cc|cc|}
\hline $D$ & $L$ & $[|\G_3|]_J/N$ & $[\bar
c(\G_3,\G_1\cup\G_2)]_J$ & $P(\G_3\not=\emptyset)$ &
$[|\G_3'|]_J/N$ & $P(\G_3'\not=\emptyset)$ & $[|\G_3\cap\G_3'|/|\G_3|]_J$ &
$P( \G_3\not=\emptyset ~{\rm and}~\G_3'\not=\emptyset )$
\\ \hline
{\bf 3}& 4  & 0.048$\pm$0.003 & 0.55$\pm$0.015 & 0.914(4) & 0.087$\pm$0.008 & 0.834(6) 
       & 0.23$\pm$0.019 &0.772(8) \\
       & 5  & 0.046$\pm$0.003 & 0.52$\pm$0.015 & 0.914(4) & 0.085$\pm$0.009 & 0.882(5) 
       & 0.15$\pm$0.016 &0.818(7) \\
       & 6  & 0.043$\pm$0.003 & 0.48$\pm$0.015 & 0.924(3) & 0.081$\pm$0.009 & 0.896(4) 
       & 0.19$\pm$0.017 &0.832(6) \\
       & 8  & 0.036$\pm$0.003 & 0.43$\pm$0.017 & 0.905(5) & 0.076$\pm$0.010 & 0.905(5) 
       & 0.16$\pm$0.019 &0.827(8) \\ \hline
{\bf 4}& 3  & 0.045$\pm$0.003 & 0.56$\pm$0.015 & 0.928(3) & 0.094$\pm$0.010 & 0.838(6) 
       & 0.25$\pm$0.020 &0.782(8) \\
       & 4  & 0.037$\pm$0.003 & 0.48$\pm$0.015 & 0.908(4) & 0.061$\pm$0.007 & 0.920(3) 
       & 0.16$\pm$0.016 &0.844(6) \\
       & 5  & 0.034$\pm$0.005 & 0.43$\pm$0.024 & 0.84(1)  & 0.072$\pm$0.014 & 0.865(8) 
       & 0.19$\pm$0.027 &0.73(1) \\
\hline
\end{tabular}
\caption{The size of the spin
domain $\G_3$ and $\G_3'$, the correlation of $\G_3$ with
$\G_1\cup\G_2$ and the relative part of
$\G_3$ and of $\G_3'$, which is common to both these spin domains. 
All results are taken for realizations where the
domains concerned do not vanish, and we give also the probability of
this to happen. All data was taken for $T=0.2$. We present the
average over these realizations $\{J\}$ $\pm$ the statistical
error, obtained by dividing the standard deviation by $\sqrt{N_s}$,
where $N_s$ is the number of realizations that contributed to each
average.} \label{tab:g3}
\end{table}


\ssection{Spin space structure}
\label{sec:spincluster}
So far we have obtained the spin domains using the results of the state
space analysis.
However,
the existence of these domains can also be observed directly in spin space,
i.e.  without utilizing information about the previously identified 
hierarchical structure of state space, as we now demonstrate.

As described in Sec. \ref{sec:states}, the equilibrium ensemble of states,
obtained for each realization, is represented by an $N \times M$
data matrix $\{S^\mu_i\}$ (e.g. Fig \ref{fig:datmat}(a)). In Sec.
\ref{sec:states} we treated each of the $M$ {\it states}, represented by a
column of this matrix,  as a ``data point"  whose coordinates are
the components of this $N$-dimensional vector. Now we view each
of the $N$ {\it spins} of the system as a data point, represented by a row of
the same matrix. Each of these data points is a vector in an $M$-dimensional
space.

\begin{figure}[t]
\centerline{\psfig{figure=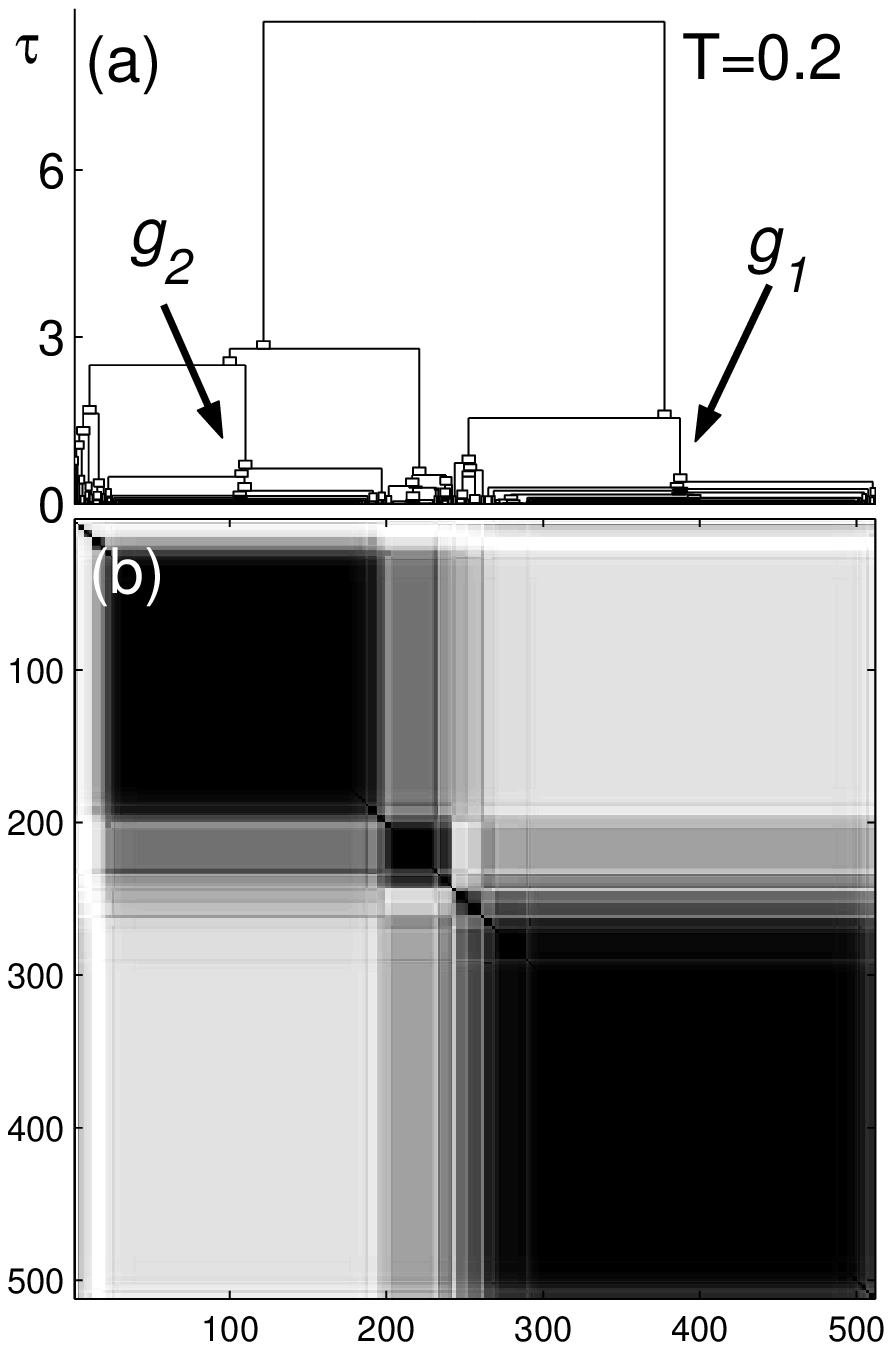,width=6cm}
\psfig{figure=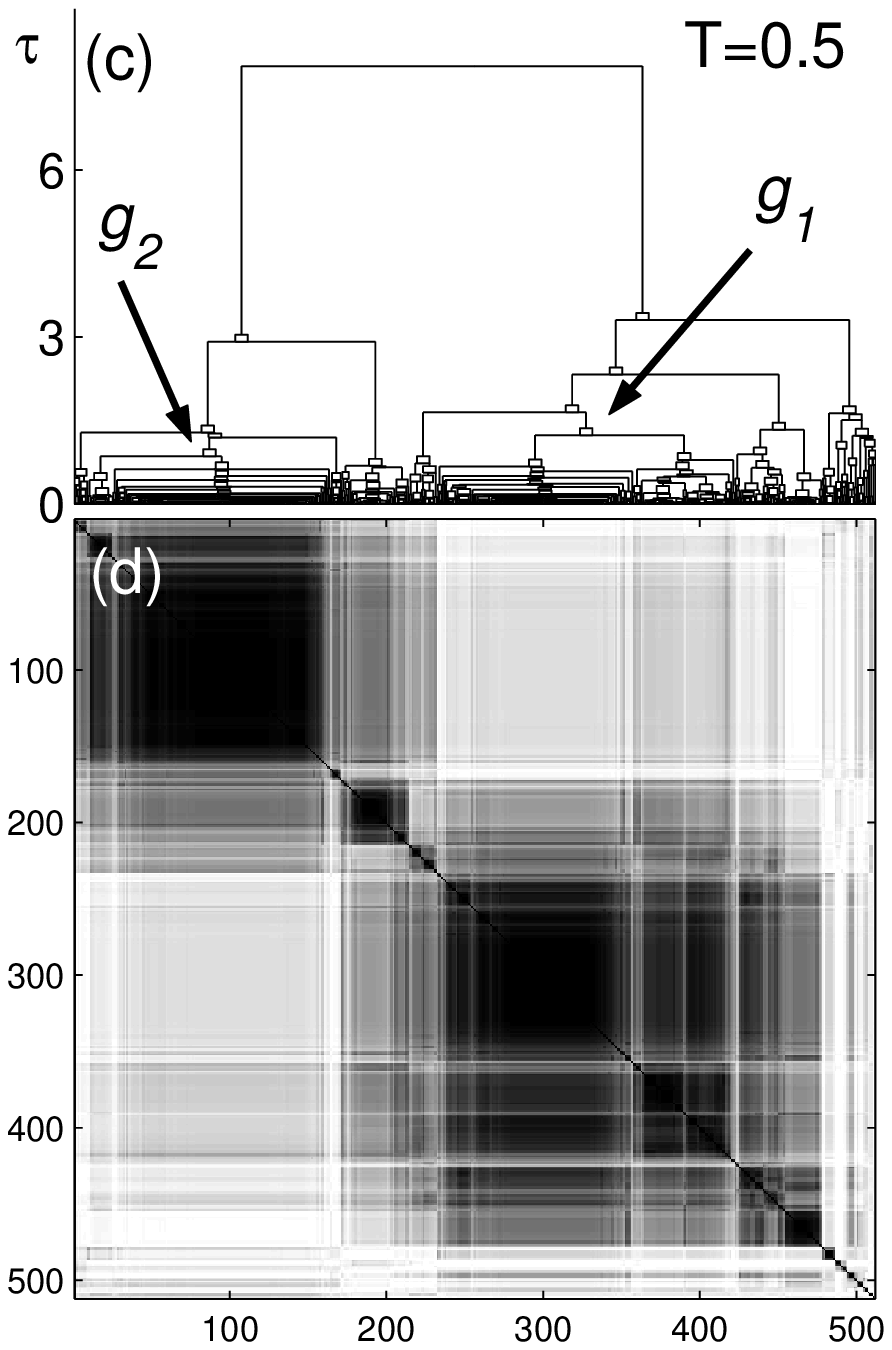,width=6cm}
\psfig{figure=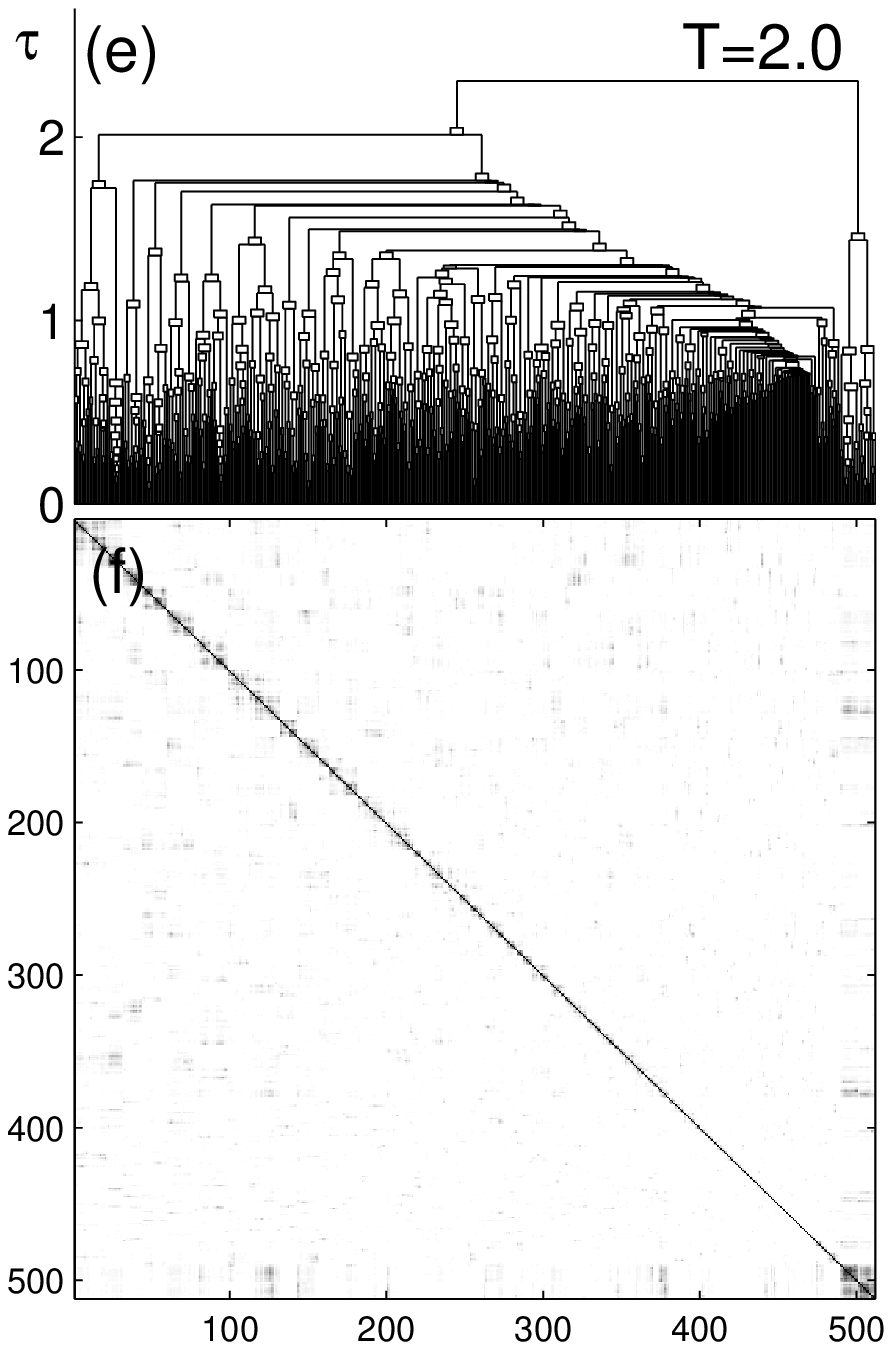,width=6cm}}
\vspace{2mm}
\caption{
{\bf (a)} The spin dendrogram $\cal D$ for the data of Fig.
\ref{fig:datmat}(a) produced by Ward's algorithm.
{\bf (b)} The spin distance matrix $d$ of this realization realization.
The spins are ordered according to their clusters in $\cal D$.
Darker shades correspond to smaller distances and higher correlations.
{\bf (c), (d)} The same as in (a), (b), for the same realization at $T=0.5$.
{\bf (e), (f)} The same as in (a), (b), for the same realization at $T=2.0$.
The $y$-axis is rescaled to show the dendrogram, which
clearly differ from the dendrograms in (a) and (c). }
\label{fig:spin_dend}
\end{figure}

The distance on the set of spins should be defined according to the nature
of the clusters we are interested in. At this case, we expect highly correlated
spins to be in the same cluster, and spins with low correlation to be in
different clusters. Thus, we define the distance between a pair of data
points $i$ and $j$ as
\beq
d_{ij}=1-{c_{ij}}^2 \;.
\label{eq:dij}
\eeq
This $N \times N$ distance matrix serves as the input for
clustering the spins, using Ward's algorithm. The dendrogram $\cal D$,
obtained when
the data of Fig. \ref{fig:datmat}(a) are clustered,
is presented in Fig. \ref{fig:spin_dend}(a). The correlated spin clusters 
are represented by boxes in the dendrogram - let us denote them by $\tilde g_a$.
When the spins are
reordered according to the dendrogram, their distance
matrix, shown in \ref{fig:spin_dend}(b), clearly exhibits a
non-trivial structure. There are large, highly
correlated spin clusters on the lower
levels of the dendrogram.

\begin{figure}[ht]
\centerline{\psfig{figure=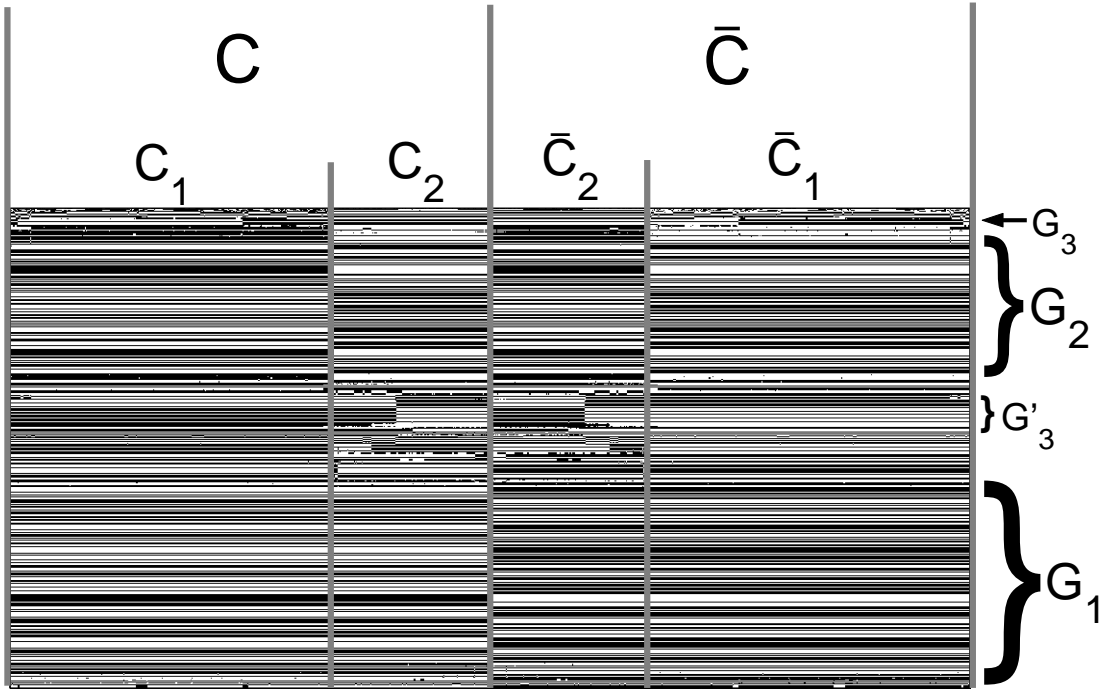,width=15cm}}
\caption{A redrawing of the ordered data matrix of
Fig. \ref{fig:datmat}(c), in order to highlight the state clusters and spin
domains discussed in the text.
It is for a $3 D$ realization of size $N=8^3$ at $T=0.2$.
The columns represent the states $\mu$ and the
rows represent the spins $i$, $S^\mu_i=\pm1$, with black/white
representing $+/-$. The states are ordered according to the
dendrogram in Fig. \ref{fig:state_dend}, and the spins ordered according
to the spin dendrogram $\cal D$ in Fig. \ref{fig:spin_dend}.
The state clusters and the spin domains are marked (see text).
}
\label{fig:datmatc}
\end{figure}

In order to ``see" the manner in which the spins are ordered, we
return to the data matrix of Fig. \ref{fig:datmat}(a). We obtained
Fig. \ref{fig:datmat}(b) from (a) by reordering the columns
according to the state dendrogram in Fig.~\ref{fig:state_dend}. If
we now reorder the rows of Fig. \ref{fig:datmat}(b) according to
the spin dendrogram $\cal D$ in Fig.~\ref{fig:spin_dend}, we get
Fig.~\ref{fig:datmat}(c), which is redrawn as Fig.
\ref{fig:datmatc} with labeling of the largest state clusters and
spin domains. The cluster structure of the spins can be clearly be
seen in  Fig. \ref{fig:datmatc}.
Spins in $\G_1$ clearly have the same
orientation in the states of $\C$ but are inverted in the
corresponding states of $\bar\C$. Spins in $\G_2$ have opposite
orientations in $\C_1$ and $\C_2$ and are inverted in the
corresponding states of $\bar\C_1$ and $\bar\C_2$. One can also
see that spins in domain $\G_3'$ separate $\C_2$ into two
sub-clusters.
As to $\G_3$, we point
in Fig. \ref{fig:datmatc} to a few (3 - 4) spins, which have the
same sign in all states of $\C_2$ but change sign in $\C_1$.

These data were obtained at $T=0.2\ (< T_c)$.
Above $T_c$ the correlation between any two spins is low, and there
is no cluster structure, as evident from Fig. \ref{fig:spin_dend}(e,f).
The relative $\tau$ values of this dendrogram are much smaller then
those of the dendrograms in Figs. \ref{fig:spin_dend}(a,c), and the reordered
distance matrix is structure-less.
If the domains $\G_a$ (that were identified in Sec. \ref{sec:spindomains} 
on the basis of the state
hierarchy) are not an artifact of our
analysis, they should be clearly identifiable in spin space, and appear
as clusters in the spin dendrogram $\cal D$. To check this, 
for 
each realization 
we compared every spin cluster $\tilde g_a$, that appears in
the corresponding spin dendrogram $\cal D$, 
to every spin domain $\G_a$ that was previously found for that realization.
The spin cluster $\tilde g_a$ that was found to be  
most similar to $\G_a$ was identified and denoted by $g_a$. 
We used the similarity measure
\beq
{\cal S}(g_a,\G_a) = { 2 |g_a \cap \G_a| \over |g_a| + |\G_a| }
\label{eq:spinsim}
\eeq
which represents the fraction of shared spins by the ``physical spin domain"
$\G_a$ and the spin cluster $g_a$.
For most realizations we have (at low $T$) $g_a=\G_a$ for both $a=1,2$;
and when these groups are not precisely equal, they
differ by only a few spins (see~\cite{HedThesis} for full details).

Fig. \ref{fig:datmatc} also provides a convenient,
simple ``geometrical" interpretation
of the two tests for the survival of our picture in the large $L$
limit that we discussed in Sec. \ref{sec:spindomains}.
Observe the
rectangular region corresponding to spin domain $\G_2$ and state
cluster $\C_2$. Validity of our picture relies on ``survival" of
this rectangle as we take the $L \rightarrow \infty$ limit. The
first test we performed checked whether its vertical side, $|\G_2
|$ stays finite. If this condition is not satisfied, the relative
area of our  rectangle goes to zero; a non-vanishing limiting
$|\G_2|$ does not, however, guarantee that the rectangle stays
finite; it may disappear if its horizontal dimension shrinks to
zero when $L \rightarrow \infty$. The second test, showing that
the correlation ${\bar c}_{12}$ does not approach 1, 
ensures that this does not happen either.

Overall, Fig. \ref{fig:datmatc} summarizes in a convenient pictorial way our
picture of the spin glass state in short range systems.

\ssection{Spin domains and states hierarchy}
\label{ssec:spin_hr}
Now that the spin domains have been well defined, we can examine the manner in
which they govern the hierarchical partitioning of state space.
Each state cluster at level $a$ of the hierarchy
can now be identified with one of two possible configurations of the particular
spin domain $\G_a$. We denote these two  configurations
as $\Uparrow_a$ and $\Downarrow_a$. Note that we have
avoided the notation $+/-$ for the states of the spin domains,
since in each state some of the spins have the $+$ sign and  others $-$.
For example, in the first level partition $\G_1$ has a certain characteristic
configuration, $\Uparrow_1$, over all the states in $\C$,
whereas over all the states of $\bar\C$ it is
in the spin inverted configuration $\Downarrow_1$.
The value $[\Uparrow_1]_i$, taken by spin $i\in\G_1$ in the
configuration $\Uparrow_1$, is defined by
\beq
[\Uparrow_1]_i = {\rm sign}\left( \sum_{\mu\in\C} S^{\mu}_i \right) \;.
\eeq
Our definition of $\G_1$,
using Eq. (\ref{eq:tg}) with $\theta=0.95$,
guarantees that the argument of the sign function
in the above expression does not vanish.
Hence, stating that $\G_1$ takes configuration $\Uparrow_1$ in a certain state
$\mu$ implies that
\beq
\sum_{i\in\G_1} S^{\mu}_i [\Uparrow_1]_i > 0 \;.
\eeq
The configuration assumed by $\G_1$ in any state
$\mu$ determines that $\mu$ is assigned to $\C$ if $\G_1$
is in configuration $\Uparrow_1$, or to $\bar\C$ if $\G_1$ is in
configuration $\Downarrow_1$.

The spin domain $\G_2$ determines, in a similar way, the partition of $\C$
into $\C_1$ and $\C_2$ (and the partition of $\bar\C$ into $\bar\C_1$ and
$\bar\C_2$). $\G_2$ is in
configuration $\Uparrow_2$ in states $\C_1$
and $\bar\C_2$, and in $\Downarrow_2$
in states $\C_2$ and $\bar\C_1$ (see Fig. \ref{fig:scheme} for a schematic
illustration of this point).

Each spin domain $\G_a$ defines a partition of the states, at
level $a$, into two groups - one in which $\G_a$ is in the
$\Uparrow_a$ configuration and the other with $\Downarrow_a$.
Picking a pair of states $\mu$ and $\nu$, one from each group, the set
of spins $\G_{\mu\nu}$, that are flipped in the transition between them,
will always include $\G_a$ ~\cite{foot3}.
Thus, the
distance $D_{\mu\nu}=|\G_{\mu\nu}|/N$ between two such states will almost
always be larger then $|\G_a|/N$.

By our definition of $\G_a$, the probability that a large part
of its spins will lose their relative orientation is small.
Considering local dynamics, the time it will take $\G_a$ to flip
is exponential in its size. If $\G_a$ is macroscopic (as we have
shown for $a=1,2$) it may be associated with a macroscopic free
energy barrier. In an infinite system it will take an infinite time
to flip, thus inducing a separation of the phase space into two
ergodic sub-spaces (or valleys).

The clear hierarchical organization of the state clusters suggest
that the average distance (\ref{eq:dcc}) between state clusters
formed at a high level of the hierarchy is significantly larger then the
average distance between clusters formed at a lower level. Indeed, we show
in Table \ref{tab:dcc} that in general $D(\C,\bar\C) \gg
D(\C_1,\C_2)$. We relate this characteristic of the state
structure to the large variability of the spin domain sizes
$|\G_a|$. Indeed, we have seen that typically  $|\G_1| > 8|\G_2|$
for $T=0.2$, $D=3,4$.

Now we have a complete picture, supported by our numerical findings, of
a hierarchy of state clusters.
The valleys are the leaves of this hierarchy ~\cite{foot4}.
At each level $a$ of this hierarchy the
partition of the states is refined according to the orientation of
macroscopic spin domains $\G_a$. At different nodes
of a certain level of the hierarchy there might be different correlated
domains that determine their partition. Take, for example,
the states in $\C_1$ (where $\G_1$ is in configuration $\Uparrow_1$
and $\G_2$ is in configuration $\Uparrow_2$). Over these states
the largest {\it unlocked}~\cite{foot5}
correlated domain is $\G_3 = \G_3(\Uparrow_1,\Uparrow_2)$.
The two possible configurations of $\G_3$ inside
$\C_1$ may be denoted as $\Uparrow_3\!(\Uparrow_1,\Uparrow_2)$ and
$\Downarrow_3\!(\Uparrow_1,\Uparrow_2)$.
Over the states of $\C_2$ we expect to find a different unlocked correlated
domain  $\G_3' = \G_3'(\Uparrow_1,\Downarrow_2)$.
We calculated the part of each domain which is included in the other.
The results are given in Table \ref{tab:g3}. We see that $\G_3$ and
$\G_3'$ share in general less than a fifth of their spins.

Note that in the ideal case (corresponding to $\theta = 1$), a spin domain
$\G_a (\Uparrow_1,\Uparrow_2,...\Uparrow_k)$,
that appears at a particular level of the hierarchy, cannot  share spins
with the higher level domains $b=1,2,...,k$, whose orientation is fixed while 
$\G_a$ flips.
For $\theta = 0.95$ such sharing was also practically excluded. On the other
hand, two domains such as $\G_3$ and $\G_3'$ {\it can} have shared spins, namely
those that are free to flip in both the $(\Uparrow_1,\Uparrow_2)$ and
$(\Uparrow_1,\Downarrow_2)$ situations.

Going all the way down the states hierarchy, we find that each valley
can be characterized by a specific list of domain configurations, e.g.
$\{ \Uparrow_1,\;\Downarrow_2,\;\Downarrow_3\!(\Uparrow_1,\Downarrow_2),
\;\Uparrow_4\!(\Uparrow_1,\Downarrow_2,\Downarrow_3),\;\ldots\;\}$.


\begin{figure}[ht]
\centerline{\psfig{figure=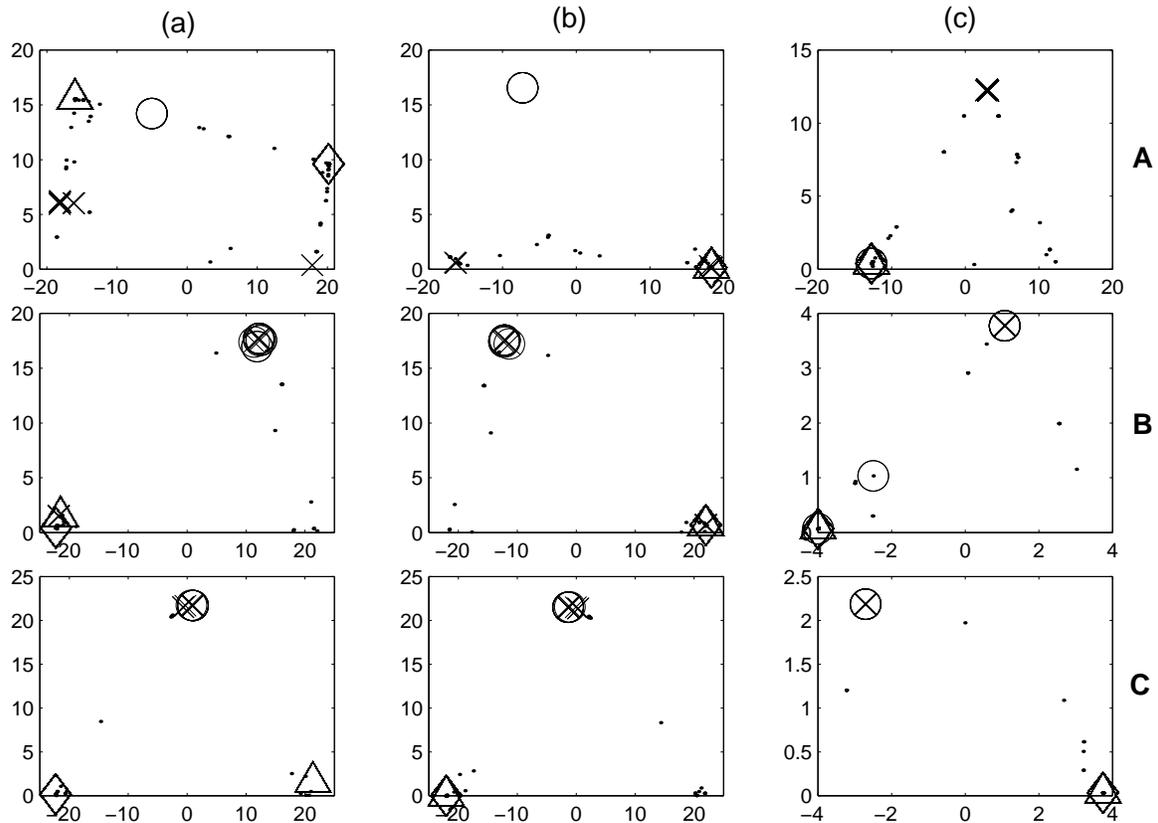,width=15.5cm}}
\vspace{2mm}
\caption{The two principal components of the 512 spins of three realizations
A, B and C (see text) in $3 D$. Each point represents a spin $i$ and its
coordinates are the projections of ${\mathbf S}_i = (S_i^1, S_i^2,
\cdots,S_i^M)$ on to the two largest eigenvectors of the matrix $R$ in
Eq.~(\ref{eq:Rspin}).
The analysis is carried over (a) all states; (b) the states of $\C_1$; and
(c) the states of $\C_2$. The spins of $\G_1$ are marked by $\Diamond$;
of $\G_2$ by $\triangle$; of $\G_3$ by $\bigcirc$; and of $\G_3'$ by $\times$.
Spins that belong to both $\G_3$ and $\G_3'$ are marked by $\bigotimes$.
Spins that do not belong to any of these domains are
marked with dots. The lower half of the plane is projected onto the
upper half using $(x,y)\rightarrow(-x,-y)$.
Spins in a correlated domain usually have the same
values for the two principal components, and they fall on top of
each other on the plot. Therefore, in most plots, a correlated domain
seems to be represented by a single marker.}
\label{fig:spinpca}
\end{figure}

An additional insight is obtained from a PCA of the spins, which
is to be distinguished from the PCA of the states in
Fig.~\ref{fig:pca}. To perform the PCA of the spins we form the
covariance matrix
\begin{equation}
R_{\mu\nu} = {1\over N} \sum_{i=1}^N \delta S_i^\mu \delta S_i^\nu \;,
\label{eq:Rspin}
\end{equation}
which is analogous to Eq.~(\ref{eq:Rstate}),
and project the two largest eigenvectors of
$R$ on to the spin configurations ${\mathbf S}_i$ for each site $i$.

The results for three realizations, labeled A, B and C, are shown in Fig.
\ref{fig:spinpca}.  Each data point represents one spin.
Realization A is the one whose data matrix is shown in
in Figs. \ref{fig:datmat}(c) and
Fig. \ref{fig:datmatc}.

In the upper left frame of Fig. \ref{fig:spinpca} we see the results
of the PCA analysis of the spins
for realization A.
We want highly correlated spins to be close on
the plot. Since a spin ${\mathbf S}_i$ is fully correlated with its
inverse $-{\mathbf S}_i$
each point $(x,y)$ with $y<0$ is projected on the plot to $(-x,-y)$.
The spins of $\G_1$ are highly correlated with each other and all have
the same values for the first two principal components of the spin space.
Therefore they fall on top of each other, and we see only one $\Diamond$
marker which represents all of them. The same is true for the spins of $\G_2$,
marked by $\triangle$. As seen from  Fig. \ref{fig:datmatc} the spins of
$\G_1$ are not correlated with the spins of $\G_2$ over the $M$ states,
and indeed the two domains are far from each other on the plot.

In column (b) of Fig. \ref{fig:spinpca} we used only the states of
$\C_1$ in the analysis. We can see in Fig. \ref{fig:datmatc} that over
$\C_1$ the spins of $\G_1$ and $\G_2$ are correlated, together with
some of the spins of $\G_3'$, marked by $\times$.
In the plot (the middle frame on the
upper row of Fig. \ref{fig:spinpca}) we can see that indeed these
spins are all plotted at the same coordinates. The spins of $\G_3$,
marked as $\bigcirc$, are highly correlated, but are not correlated with
$\G_1$ and $\G_2$. Note that the spins of $\G_3'$ are separated into two
different sets, and are not correlated over $\C_1$.

When we perform the analysis using only the states of $\C_2$ we get the
results presented in column (c) of Fig. \ref{fig:spinpca}. In the
matrix of Fig. \ref{fig:datmatc} we see that the spins of $\G_1$, $\G_2$ and
$\G_3$ are correlated together over $\C_2$, and indeed they all fall
on top of each other in the plot. We also see $\G_3'$ as a separated
correlated domain.

In the second row of Fig. \ref{fig:spinpca} we give the results for
realization B, in which $\G_3$ and $\G_3'$ share some of their spins.
Those spins are marked by $\bigotimes$. In column (c) we see these
spins inside $\G_3'$. The rest of the spins of $\G_3$ are not correlated
with them. Some of them are correlated with $\G_1$ and $\G_2$, and
others seem to be in another domain.

In the third row of Fig. \ref{fig:spinpca} we present the results for
realization C in which $\G_3\subset\G_3'$. Here spins of $\G_3'$ seem
to form a correlated set also over $\C_1$, though the correlations
are not high enough for it to be considered as a domain by our definition.

\section{State overlap}
\label{sec:overlap}


We have presented a description of the system in its low $T$
phase, relating state space behavior to the microscopic structure
in spin space.
Most of the previous literature, however, did not directly measure the microscopic features
of the system
but examined their indirect implications on other parameters, such
as the widely addressed overlap distribution $P(q)$. Beyond making
contact with the literature, which concentrates on measuring
$P(q)$, the aim of this Section is two fold: (i) we show how our
methods allow a useful decomposition of this function into its
physically relevant constituent parts, and (ii) we demonstrate that our
picture provides a {\it microscopic interpretation} of the
observed $P(q)$.
To this end we focus here on $P_J(q)$, the overlap distribution
for a specific realization $\{J\}$ of the bonds, whereas earlier
works \cite{Bhatt90,Marinari98,KPY00,Hartmann3d} presented results
for the average over the disorder, $P(q)=[P_J(q)]_J$.

Two technical comments should be first made. First, because of
overall spin-flip symmetry, the function $P_J(q)$ is symmetric and
hence we can limit our attention to $q > 0$. Second, since for
most realizations $|\G_1|>N/2$, we have
\beq
 P_J^{\C \C}(q) \simeq \left\{
\begin{array}{ll}
P_J(q) & q\geq0 \\
0 ~~~~~ & q<0 \\
\end{array}
\right.
\eeq
where by $P_J^{\C \C}(q)$
we denote the distribution of
overlaps between pairs of states $\mu,\nu \in \C$, so that we have
to deal only with such pairs.

\subsection{Decomposition of $P_J(q)$ and $P(q)$}

The overlap distribution for a specific realizations of the
randomness, $P_J(q)$, is expected to be the sum of two main parts
\begin{equation}
P_J(q) = P^i_J(q) + P^o_J(q),
\label{eq:PiPo}
\end{equation}
where $P^i_J(q)$ is the overlap distribution {\it within} a valley
(and between a valley and its spin reversed
counterpart), and $P^o_J(q)$ is the overlap distribution between
states that belong to {\it two different} valleys.  $P^i_J(q)$
converges to $\delta(|q|-q_{\rm EA})/2$ in the thermodynamic
limit, where $q_{\rm EA}$ is the Edwards-Anderson order parameter,
which will also be denoted as the ``self-overlap''.  $P^o_J(q)$ is
the sum of several contributions, corresponding to different pairs
of valleys.

In the thermodynamic limit this separation is unambiguous; if two
microstates $\mu$ and $\nu$ are separated by a macroscopic
energy barrier, they belong to two different valleys and their
overlap $q_{\mu \nu}$ contributes to $P^o_J(q)$. For finite
systems this separation is problematic; our picture and method,
however, does allow us to estimate $P^o_J(q)$ or, to be more
precise, to calculate a function ${\tilde P}^o_J(q)$ defined
below, which is a {\it lower bound} to it. In our picture, the
transition between such pairs of microstates (that belong to two
different valleys) is associated with flipping a specific set
of spin domains. Consequently, having identified the relevant spin
domains, we can identify when $\mu$ and $\nu$ belong to different
valleys and also the level in the states' hierarchy at which
they differ.

A remaining apparent ambiguity concerns the level of the state
hierarchy at which we ``stop'' and decide whether a particular pair
of microstates belongs to different valleys or not. Suppose we
stop  the decomposition of $\C$ at some level $n$ and 
denote by $\C_\alpha^n$ the clusters
obtained at this level. The overlaps obtained from pairs of
microstates that belong to different valleys {\it at this
level} are assigned to the distribution $P_J^{o,n}(q)$, and pairs
from the same valleys to $P_J^{i,n}(q)$:
\begin{eqnarray}
P^{i,n}_J(q) & = & \sum_\alpha P_J^{C_\alpha^n C_\alpha^n}(q) \nonumber \\
P^{o,n}_J(q) & = & \sum_{\alpha \ne \beta} P_J^{C_\alpha^n
C_\beta^n}(q) , \label{eq:pi_po}
\end{eqnarray}
where, from Eq.~(\ref{eq:PiPo})
\begin{equation}
P_J(q) = P^{i,n}_J(q) + P^{o,n}_J(q) \qquad {\mathrm for }\ q \ge
0 .
\label{eq:pippo}
\end{equation}

Clearly, by going down a level further, to $n+1$, some pairs that
were assigned to $P_J^{i,n}(q)$ will be reassigned to
$P_J^{o,n+1}(q)$, but if a pair was in $P_J^{o,n}(q)$ it will stay
in $P_J^{o,n+1}(q)$. This argument clearly shows that
$P_J^{o,n}(q)$ obtained at any level is a {\it lower bound} to
$P^o_J(q)$. This point is explained again below for the particular
case of $n=2$.

\begin{figure}[t]
\centerline{\psfig{figure=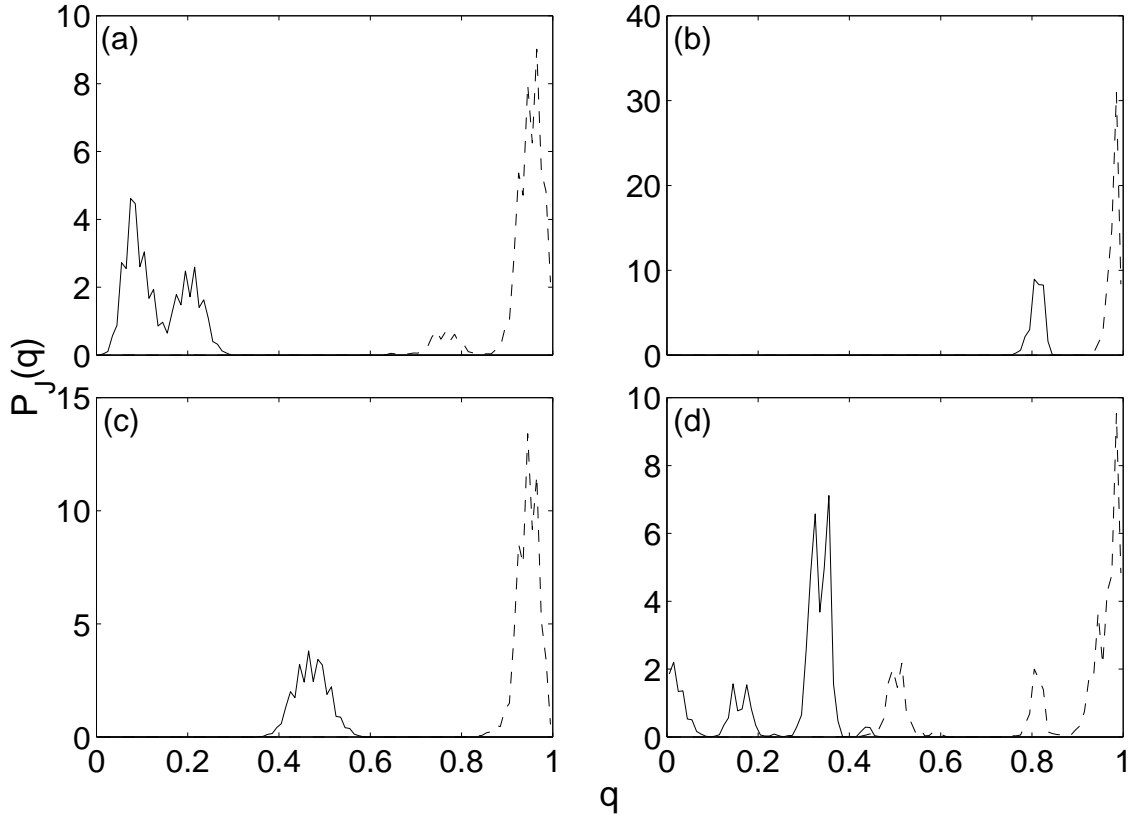,width=15cm}}\vspace{2mm}
\caption{The distribution $P_J(q)$ for four realizations of
$\{J\}$ at $T=0.2$ in $3 D$. The distribution in (a) is the same
as in the top frame of Fig. \ref{fig:pccq}. The solid line describes
$P^{C_1C_2}_J(q)$ 
and
the dashed line plots the rest of the 
distribution, $P_J(q)-
P^{C_1C_2}_J(q)$. 
The latter contains a large peak at $q \approx 1$
which is the
distribution $P^i_J(q)$, of overlaps inside the valleys.
}
\label{fig:pq}
\end{figure}

To demonstrate how natural is the separation of eq.
(\ref{eq:PiPo}), we consider pairs of states $\mu\in\C_1$ and
$\nu\in\C_2$, i.e. pairs taken from state clusters that appear at
the second ($n=2$) level of the states' hierarchy. According to
our picture such pairs contribute a non-vanishing part of
$P_J(q)$, which we denote by 
$P^{C_1C_2}_J(q)$ ($\equiv P_J^{o,2}(q)$, since for $n=2$ $\C$ has only these 
two sub clusters) 
This function, as
well as its complement $P_J(q) - P^{C_1C_2}_J(q)$ are presented,
for $T=0.2 $ and $L=8$ in Fig. \ref{fig:pq}, for four realizations
of the randomness. The figure shows clearly that the separation is
natural, and not just an artifact of our analysis.

For all these four realizations the spin domain $\G_2$ is clearly
identifiable and is ``macroscopic" (note that this holds for more
than $80 \%$ of the realizations, see table \ref{tab:c12}).
In all these cases the states $\mu$ and $\nu$ belong to
different valleys, and contribute to $P^o_J(q)$. There may be,
however, pairs of states which also contribute to $P^o_J(q)$, but
are {\it not} included in $P^{C_1C_2}_J(q)$. This happens when (at
least) one of the state clusters $\C_1, \C_2$ has internal
structure and decomposes into sub-clusters (i.e. higher level valleys).
 Say $\C_1$ contains two such sub-clusters, $\C_{1a},
\C_{1b}$. The overlap of a pair of states $\mu \in \C_{1a}$ and
$\nu \in \C_{1b}$ contributes to $P^o_J(q)$, and is not included
in $P^{C_1C_2}_J(q)$; hence the latter function is a {\it lower
bound} on the former.
As discussed above in Sec.~\ref{sec:spins}, such internal structure of $\C_1$
(or $\C_2$) is associated with a spin domain $\G_3$ (or
$\G_3^\prime$). 
This structre is clearly present for the
realizations in Fig. \ref{fig:pq} (a) and (d), as evident from the
multi-peaked structure of $P^{C_1C_2}_J(q)$s which is discussed further below.

\begin{figure}[t]
\centerline{
\psfig{figure=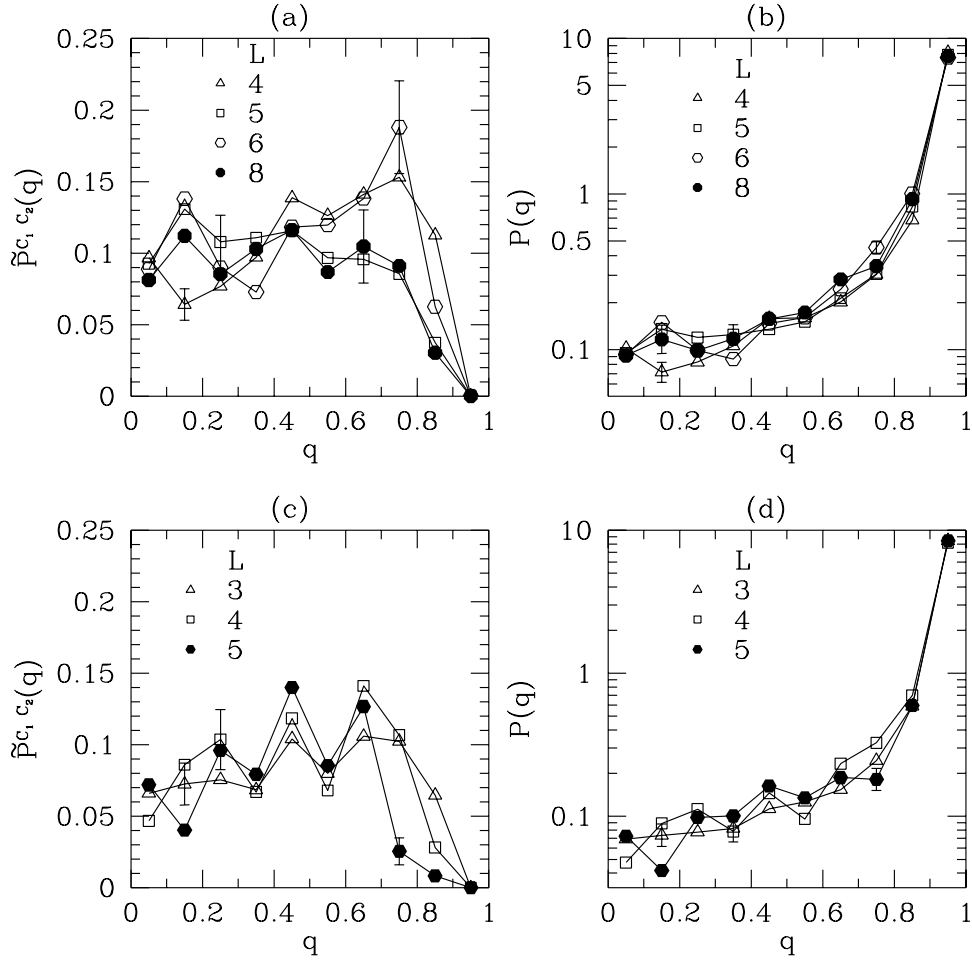,height= 13cm}
} \caption{
{\bf (a)} The
partial distribution $\widetilde P^{C_1C_2}(q)$ for
$D=3~L=4,5,6,8$. It is normalized so that $2\int_0^1
\widetilde P^{C_1C_2}(q)$ is its weight in the total $P(q)$. For clarity only
a few representative error bars are shown.
{\bf (b)} The distribution $P(q)$ for
the same systems as in (a).
{\bf (c)} $\widetilde P^{C_1C_2}(q)$ as in (a) but for
$D=4~L=3,4,5$.
{\bf (d)} $P(q)$ for the same systems as in (c).
}
\label{fig:pqo}
\end{figure}

We now generate a distribution $\widetilde P^{C_1C_2}(q)$ which is a lower
bound on the contribution of $P^{C_1C_2}(q) \equiv
[P^{C_1C_2}_J(q)]_J$ to the {\em average} distribution $P(q)$. In
order to assure that $\widetilde P^{C_1C_2}(q)$ constitutes a
lower bound to $P^{C_1C_2}(q)$, we included in $\widetilde P^{C_1C_2}(q)$ 
{\em only}\/ contributions $P^{C_1C_2}_J(q)$
from those realizations $J$ in which $\G_2$
was relatively large, namely $|\G_2|>0.05N$. For the other 
realizations we
set the contribution to the average over $J$ to zero; hence our
$\widetilde P^{C_1C_2}(q)$ is a lower bound to the true
$P^{C_1C_2}(q)$ (which, in turn, is a lower bound to $P^o(q)$). In
Figs. \ref{fig:pqo} we show the distributions $P(q)$ and
$\widetilde P^{C_1C_2}(q)$. The data indicates that the weight in
the tail for small $q$ stays finite with increasing $L$ (at least
for this range of sizes), in agreement with earlier studies
\cite{Bhatt90,Marinari98,KPY00,Hartmann3d} which just measured
$P(q)$. For systems with Gaussian couplings $P^i(q)$ has a very
small contribution at $|q|<0.7$ and $\widetilde P^o(q)$ is the
dominant part of $P(q)$ in this range.
For an Ising spin-glass with binary couplings, however, the
difference between the distributions is significant and proper
care must be taken when delicate issues, such as triviality of
$P(q)$, are investigated~\cite{HedEPL}.

\subsection{Interpretation of $P_J(q)$ in terms of spin domains}
\label{sec:spindomns}
Our aim is to interpret the distribution $P_J(q)$, obtained for a
particular realization,  in terms of the state clusters $\C_i$ and
spin domains $\G_a$ that were discussed in the previous sections.
Before going into a detailed discussion and analysis, we state the
interpretation that arises, for the four realizations whose
$P_J(q)$ was shown in Fig. \ref{fig:pq}. The first of these, Fig.
\ref{fig:pq} (a), corresponds to a system in which $\C_2$ has
internal structure, due to a sizeable domain $\G_3^\prime$; its
counterpart, $\G_3$ is too small to have a clear signature. The
size of $\G_3^\prime$ governs the splitting of the peak drawn with
a solid line and also of the peak at high $q$ (dashed line). In
the systems of Fig. \ref{fig:pq} (b) and (c) neither  $\C_1$ nor
$\C_2$ have noticeable internal structure; the domains
$\G_3,\G_3^\prime$ are microscopic. The system of \ref{fig:pq} (d)
has internal structure for both $\C_1$ and $\C_2$, induced by
domains $\G_3$ and $\G_3^\prime$, respectively. The sizes of these
two domains govern the observed splitting of both the solid and
dashed curves.

One can  associate each peak of $P_J(q)$ with the overlaps of pairs
of states that are related by flipping one or more of the
previously identified spin domains. In this regard our
interpretation resembles the RSB picture~\cite{RSB} which also
relates the peaks of $P(q)$~\cite{Marinari98}
to overlaps between configurations in different valleys.

\begin{figure}[t]
\centerline{\psfig{figure=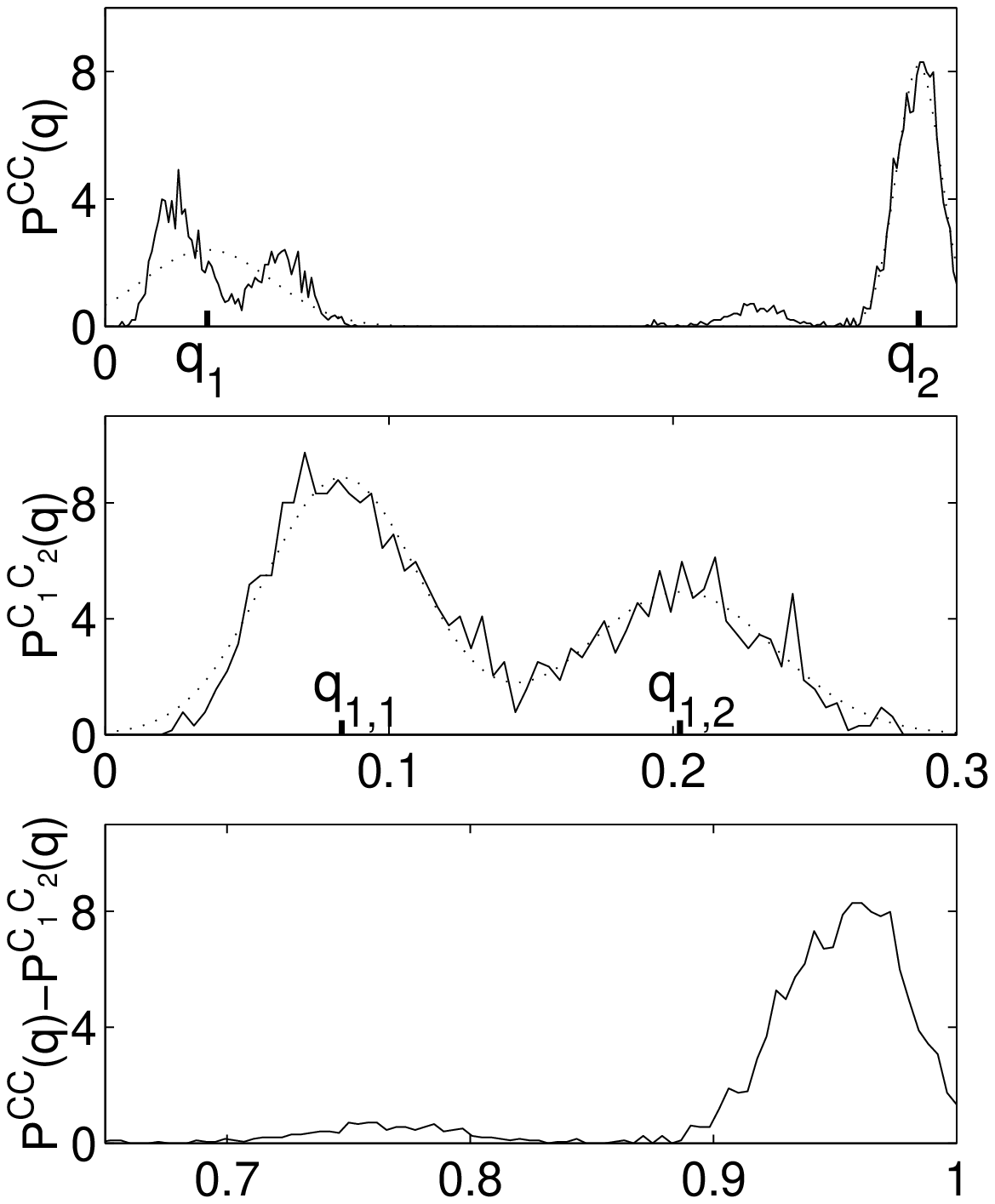,width=10cm}}\vspace{2mm}
\caption{
(Top:) The distribution $P^{\C\C}_J(q)$ for the same
$3 D$ realization whose data were presented in Fig. \ref{fig:datmatc}.
The dotted line is a fit to the sum of two Gaussians (see text).
(Middle:) The partial distribution $ P_J^{\C_1\C_2}(q)$ for the
same realization. The dotted line represents a fit to the sum of
two Gaussians.
(Bottom:)
The difference between the two previous distributions.
}
\label{fig:pccq}
\end{figure}

To substantiate these claims and make them more precise we
consider in detail the realization whose (ordered) state and spin
data matrix is given in Fig. \ref{fig:datmatc}, and whose $P_J(q)$
(shown in Fig. \ref{fig:pq} (a)) is reproduced and magnified in
Fig. \ref{fig:pccq}. For this realization we clearly identified
{\it three} spin domains; $\G_1$, $\G_2$ and $\G_3^\prime$.
Disregarding the splitting induced by $\G_3^\prime$ (and $\G_3$,
if present) we identify two main peaks that dominate $ P_J^{\C
\C}(q)$.
We performed a fit of
$ P_J^{\C\C}(q)$ 
to a sum of two Gaussians,
\beq
P_J^{\C\C}(q)=b_1\exp[(q-q_1)^2/{a_1}^2] + b_2\exp[(q-q_2)^2/{a_2}^2] \;,
\label{eq:pqfit}
\eeq
with $a_i$, $b_i$ and $q_i$ as fit parameters, yielding the dotted
curves in the upper part of Fig. \ref{fig:pccq}.
The center of the (split) peak at
low $q$ is $q_1$ and the high-$q$ data is centered at $q_2$.

To see how these $q_i$ are related to our state clusters and spin
domains, note that the overlap $q_{\mu\nu}$ between states $\mu$
and $\nu$ is related to the size of the set $\G_{\mu\nu}$ (defined
in (\ref{eq:gmn})), of spins that flip when passing from state
$\mu$ to $\nu$;
\beq
q_{\mu\nu} = 1 - 2|\G_{\mu\nu}| / N \;.
\eeq
For nearly all state pairs $\mu,\nu \in\C$ the domain $\G_1$ is in
the state $\Uparrow_1$; hence  $|\G_{\mu\nu}| \leq 1-|\G_1|$, so
that $q_{\mu\nu} \geq 2|\G_1|/N -1$.
The state pairs belong to one of two types:
\begin{enumerate}
\item Pairs in which
$\G_2$ flips between $\Uparrow_2$ to $\Downarrow_2$ or vice versa.
These pairs contribute to $P^{o,1}_J(q)= P^{\C_1\C_2}_J(q)$. The
definition of $\G_2$ yields that in most such cases $\G_2 \in
\G_{\mu\nu}$ and hence
$2|\G_1|/N-1 \leq q_{\mu\nu} \leq 1-2|\G_2|/N$.
\item Pairs in which neither $\G_1$ nor $\G_2$ flip
contribute to $P^{i,1}_J(q)=P^{\C_1\C_1}_J(q)+P^{\C_2\C_2}_J(q)$.
For these pairs in most cases $|\G_{\mu\nu}|\leq N-|\G_1\cup\G_2|$
and hence $q_{\mu\nu} \geq 2(|\G_1|+|\G_2|)/N - 1$.
\end{enumerate}
The peak centered at $q_1$, is attributed to state pairs of the
first type, and hence

\beq
2|\G_1|/N-1 \leq q_1 \leq 1-2|\G_2|/N
\label{eq:ineq1}
\eeq
The other peak,
centered at $q_2$, is attributed to state pairs of the second
type, and thus we expect
\beq
q_2 \geq 2(|\G_1|+|\G_2|)/N - 1
\label{eq:ineq2}
\eeq
These two inequalities yield $q_2 - q_1 \geq 2|\G_2|/N$.
Evidently, this structure of $P_J(q)$ is completely consistent
with our picture of spin domains that govern partition of
state space into well defined clusters. By a detailed analysis~\cite{HedThesis}
we have shown that the (at least) two-peaked structure of $P_J(q)$ 
survives for large $L$. 


In some realizations, such as the ones that yield Figs.
\ref{fig:pq}(a) and \ref{fig:pq}(d) $P_J(q)$ has more peaks, since
$P^{\C_1\C_2}_J(q) (\equiv
P^{o,2}_J(q))$ 
exhibits two or more peaks; this splitting is
due, as mentioned above, to spin domains $\G_3$ and $\G_3'$. We
analyzed $P^{\C_1\C_2}_J(q)$
in the same way as we did for $ P_J^{\C,\C}(q)$,
using the same form of fit as in Eq.~(\ref{eq:pqfit}).
For example, in the middle part of Fig.~\ref{fig:pccq},
$\tilde q_{1,1}$ and $\tilde q_{1,2}$ denote the
centers of the two Gaussians, with $\G_2$ and $\G_3$ playing
the previous roles of $\G_1$ and $\G_2$.

For much larger systems, for which the state hierarchy is expected to have
more than two clear levels, we expect to find a finer structure in $P(q)$.  It
will exhibit multiple peaks, each related to different domain sizes. The
heights and widths of the peaks are expected to be governed by the sizes of
the state clusters that contribute to it which, in turn, are determined by the
correlations between the spin domains that generating these clusters.  Each of
these peaks can be isolated and measured separately by observing the overlap
of states of the corresponding clusters.

The shape of $P(q)$ we describe above resembles the one assumed by
RSB. It is important to re-emphasize, however, that our $P(q)$ was
obtained for finite systems; its resemblance to the form predicted
by RSB does not necessarily mean that the latter picture is the
correct one.
In fact, previous studies\cite{KM00,PY00,KPY00,Palassini99}
of the link overlap
(defined in Eq.~(\ref{eq:qlink})) indicate that it is trivial, which
contradicts the RSB scenario, though this conclusion has been disputed in
Refs.~\onlinecite{Marinari00,Marinari00b,Marinari00c}.
In fact, our picture and results
also do {\it not} appear to be consistent with RSB since we find a
non-ultrametric state structure, as we show in Sec. \ref{sec:ultm}.

\section{Ultrametricity}
\label{sec:ultm}

Ultrametricity is one of the main characteristics of the mean
field RSB picture. Efforts to establish \cite{Franz00} or dismiss
\cite{HartmannUM} the existence of ultrametricity in short range
spin glasses did not yield conclusive results.
We presented in Sec. \ref{sec:states} indications that
$w_{12}$, the width of the distance distribution between states
from $\C_1$ and $\C_2$, does not vanish, implying a
non-ultrametric structure of state space. Here we look for a more
direct test of ultrametricity. The main problem is that we can
equilibrate only small systems, where ultrametricity is hindered
by finite size effects. Ultrametricity is a statement about the
geometrical properties of {\it triangles} formed by three "pure
states" (or by three micro states that belong to different pure
states). All three have to belong~\cite{foot6}
to $\C$, and for small systems
only a small fraction of the realizations contain such triplets of
states. 

For $D=3$ at $T=0.2$ we measured $\tilde p$,  
the fraction of realizations for which
$\G_3$ (or $\G_3'$) were large enough to induce two clearly
separated peaks of $P_J^{\C_1,\C_2}(q)$
(see Sec.~\ref{sec:spindomns}). We found, for $L=4,5,6,8$ the
values ${\tilde p}=0.006,0.026,0.056,0.090$, respectively.  
At $D=4$ the similar fractions, at $T=0.2$ and for $L=3,4,5$ are 
${\tilde p}=0.02,0.030,0.080$. 
Note that for both $D=3,4$, $\tilde p$ increases with the size of the
system.

Our method of analysis allows us to identify the
realizations that do contain such triangles of states and use exclusively
them to investigate whether ultrametricity does or does not hold.
In this way we avoid many finite size effects that might obscure the
results.

A set of objects with a distance measure $D$ is ultrametric if any
three objects $\alpha$, $\beta$ and $\gamma$ form an isosceles
triangle, with the base equal to or smaller than the two equal
sides. This demand can be formulated as the requirement that the
inequality
\beq
D_{\alpha\beta} \leq \max\{ D_{\alpha\gamma},
D_{\beta\gamma} \} \;.
\eeq
be satisfied for all three choices of the distance placed on its
left side.

When the system is in the high $T$ paramagnetic phase it will
exhibit ultrametricity, since, as $L\rightarrow\infty$ the
probability distribution of distances will be
$P(D_{\mu\nu})=\delta(D_{\mu\nu}-1/2)$ and all triangles will be
equilateral. Similar behavior occurs {\it inside} a specific valley
at $T<T_c$, since for two states $\mu$ and $\nu$ inside the
valley $P(D_{\mu\nu})\rightarrow\delta(D_{\mu\nu} - 
(1-q_{\rm EA})/2)$, 
where $q_{EA}$ is the Edwards-Anderson order parameter.

The non-trivial result of RSB is that the valleys {\it
themselves} are ultrametric. In order to investigate this claim,
we have to focus on triplets of states, each chosen from a
different valley. For large systems with many valleys this
does not require special care, since almost all triplets of states
will belong to three different valleys.
For small systems, however, a large fraction of
the possible triplets will have at least two states from the same
valley. Such triplets should be disregarded.

Our way of analysis provides us with tools to examine
ultrametricity for small systems. We utilize the state hierarchy
obtained in Sec. \ref{sec:states} to carefully choose triplets of
states from different state clusters. We chose three clusters:
$\C_2$, $\C_{1a}$ and $\C_{1b}$. The last two clusters are the
``children" of $\C_1$ in the state dendrogram, i.e. $\C_1=\C_{1a}
\cup \C_{1b}$. According to our picture a triplet of states, one
from each of these three clusters, belong to three different valleys, 
since we have to flip  a correlated domain with a
macroscopic number of spins in order to move from one cluster to
another. To move from $\C_2$ to $\C_1$ we have to flip $\G_2$ from
configuration $\Downarrow_2$ to configuration $\Uparrow_2$.
Similarly, when moving from $\C_{1a}$ to $\C_{1b}$ we have to flip
$\G_3$ from $\Uparrow_3=\Uparrow_3(\Uparrow_1,\Uparrow_2)$ to
$\Downarrow_3=\Downarrow_3(\Uparrow_1,\Uparrow_2)$ (see Subsection
\ref{ssec:spin_hr}). Due to the small sizes studied,  in this
paper we do not present any conclusive evidence that $\G_3$ is
indeed macroscopic. However, if (in the $L \rightarrow \infty$
limit) it is not macroscopic, our method predicts that there are
only four valleys (determined by $\G_1$ and $\G_2$) and hence
the the RSB picture clearly does not hold.

In order to have a quantitative measure of ultrametricity we
define an index $K$ in the following manner. Let $\mu$, $\nu$ and $\rho$ be
three states, so that $D_{\mu\nu}\geq D_{\mu\rho}\geq D_{\nu\rho}$.
We define
\beq
K_{\mu\nu\rho} = {D_{\mu\nu}- D_{\mu\rho} \over D_{\nu\rho}} \;.
\eeq
The triangle inequality requires $D_{\nu\rho}\geq D_{\mu\nu}- D_{\mu\rho}$
so we have $0\leq K_{\mu\nu\rho}\leq 1$. Ultrametricity demands
$D_{\mu\nu}=D_{\mu\rho}$ so if there is ultrametricity we expect
$P(K)\rightarrow\delta(K)$ as $L\rightarrow\infty$.

We measured $P(K_{\mu\nu\rho})$ for $\mu\in\C_2$, $\nu\in\C_{1a}$
and $\rho\in\C_{1b}$. We used our samples for $T=0.2$; since as
the temperature is lower and more distant from $T_c$, the state
structure should be clearer and less blurred by finite size
effects. We measured the distribution of $K$ for each realization,
and then obtained $P(K)$ by averaging over the disorder $\{J\}$.
In all systems we found with high probability that
$K_{\mu\nu\rho}=1$ exactly (see  Table \ref{tab:pk}).This happens
when $\G_{\mu \nu}$, the set of spins one has to flip when going
from $\mu$ to $\nu$, coincides precisely with $\G_{\mu \rho} \cup
\G_{\nu \rho}$, the union of the two sets  that are flipped when
we go from $\rho$ to $\mu$ and to $\nu$. This is, however, clearly
a finite size effect; as $L$ increases the probability $P(K=1)$
decreases dramatically. Therefore we do not include this part of
the distribution in our estimation of $P(K)$. If this part of
$P(K)$
broadens
as $L$ increases, its exclusion cannot be achieved
by simply ignoring the triangles with $K=1$. This, however, is
clearly not the case: we present in Table \ref{tab:pk} the
probability $P(0.9\leq K<1)$, and show that its increase with $L$
is much too small to compensate for the decrease in $P(K=1)$.

\begin{figure}[t]
\centerline{\psfig{figure=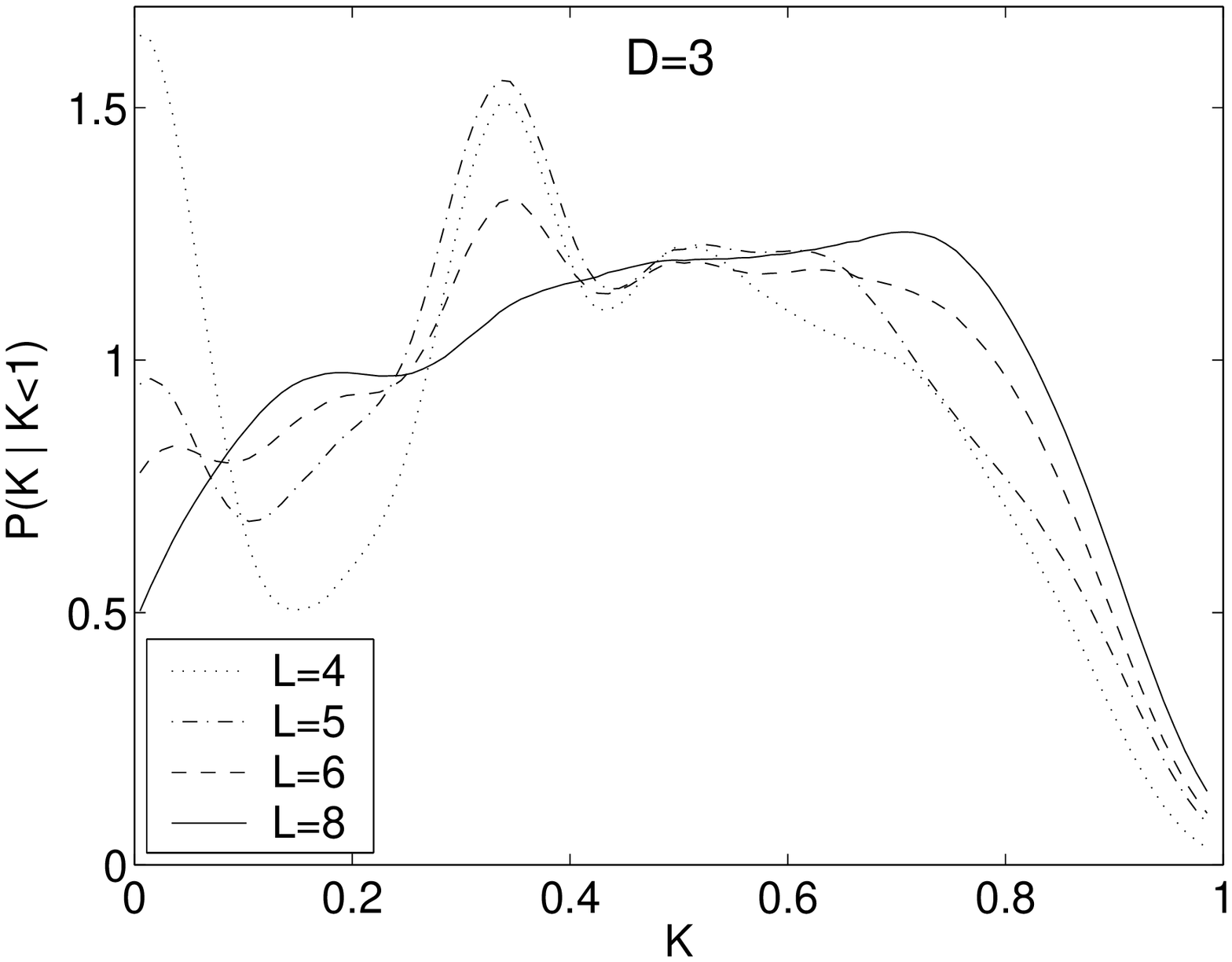,width=7cm}
~~~~\psfig{figure=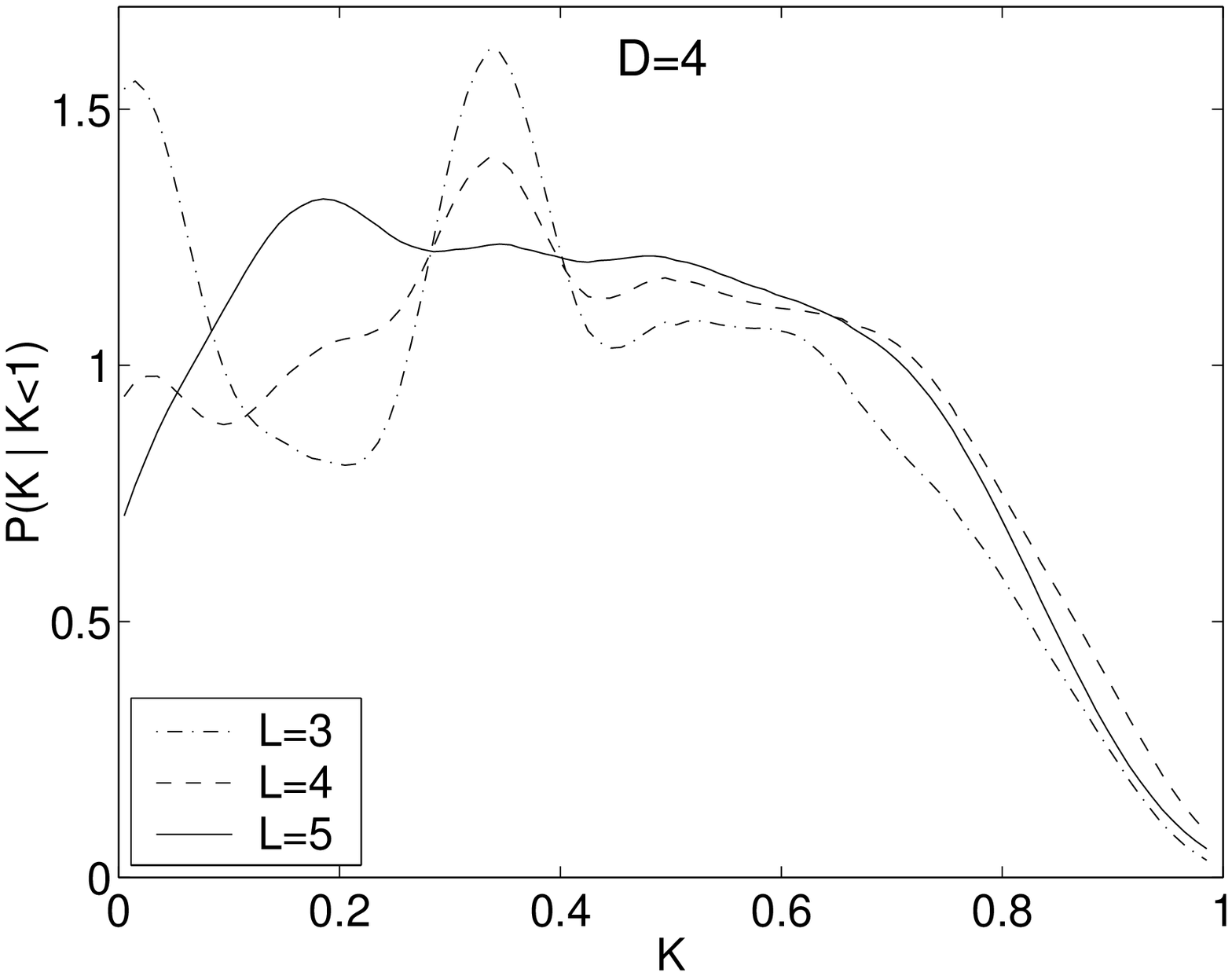,width=7cm}}
\vspace{2mm}
\caption{The distribution $P(K|K<1)$ of $K_{\mu\nu\rho}$, for $\mu\in\C_2$,
$\nu\in\C_{1a}$ and $\rho\in\C_{1b}$. All systems are sampled at $T=0.2$.}
\label{fig:pk}
\end{figure}

In order to disregard this finite size effect we truncated
$P(K=1)$ from $P(K)$ and renormalized to get the distribution
\beq
P( K | K<1 ) = \left\{
\begin{array}{ll}
P(K) \, / \, P(K<1) ~~~~ & K<1 \\
0& K=1
\end{array}
\right. \;.
\eeq
For large $L$ we expect $P(K=1)$ to vanish, and $P(K)$ will approach
$P(K|K<1)$.
The results are plotted in Fig.
\ref{fig:pk}. In Table \ref{tab:pk} we give the mean and variance of
$P(K|K<1)$. Though we deal with small systems, it seems that $P(K|K<1)$
converges to a distribution with non-vanishing mean and variance,
indicating breakdown of ultrametricity for the three valleys studied.

Again one should address the question: do these results remain
valid in the large $L$ limit? We have to show that the state
triplets we used, from $\C_2$, $\C_{1a}$ and $\C_{1b}$, have a
finite statistical weight as $L\rightarrow\infty$. In  Sec.
\ref{sec:overlap} we showed that $|\C_2|/N$ remains finite if the
average correlation $\bar c_{1 2}$ between $\G_1$ and $\G_2$ does
not approach one. From the same argument we conclude that if the
correlation of $\G_3$ with $\G_1\cup\G_2$ does not approach one
then both $|\C_{1a}|/N$ and $|\C_{1b}|/N$ do not vanish and the
weight of such state triplets remains finite, and the system does
not exhibit ultrametricity. We do have evidence that the average
correlation $\bar c(\G_3,\G_1\cup\G_2)$ of $\G_3$ with
$\G_1\cup\G_2$ in fact decreases as $L$ increases, but it is not
conclusive.

\begin{table}[p]
\begin{center}
\begin{tabular}[t]{|cc|cc|cc|}
\hline
$D$ & $L$ & $P(K=1)$ & $P(0.9\leq K<1)$ & mean$(K)$ & var$(K)$ \\ \hline
{\bf 3}&4& 0.78 & 0.0007 & 0.385 & 0.073 \\
       &5& 0.57 & 0.0082 & 0.426 & 0.066 \\
       &6& 0.35 & 0.0126 & 0.447 & 0.068 \\
       &8& 0.08 & 0.0269 & 0.476 & 0.066 \\ \hline
{\bf 4}&3& 0.74 & 0.0012 & 0.362 & 0.068 \\
       &4& 0.38 & 0.0116 & 0.413 & 0.067 \\
       &5& 0.10 & 0.0095 & 0.406 & 0.061 \\
\hline
\end{tabular}
\end{center}
\caption{The third and fourth columns show the probability for
$K_{\mu\nu\rho}=1$ and $0.9\leq K_{\mu\nu\rho}<1$, for  $\mu\in\C_2$,
$\nu\in\C_{1a}$ and $\rho\in\C_{1b}$. The fifth and sixth columns give
the mean and variance of the distribution of $P(K|K<1)$.
All systems are sampled at $T=0.2$.}
\label{tab:pk}
\end{table}

\section{Summary and discussion}
\label{sec:summary}


We have presented a new picture of the spin glass-phase in finite
dimensional systems. This picture - State Hierarchy Induced by Correlated Spin
domains (SHICS) -  is consistent with
numerical
findings of a non-trivial overlap distribution
\cite{Bhatt90,Marinari98,KPY00} and macroscopic spin domains
which cost only a {\em finite}\/ energy to flip \cite{KM00,PY00}.
Our results differ from the conventional interpretations 
\cite{Bray86,Moore98}
of the droplet picture; nevertheless, the scenario
presented in the original work of Fisher and
Huse\cite{droplet,Huse87,Fisher87},
and also the work of Newman and Stein~\cite{Newman98,Newman00},
is of sufficient generality to allow consistency with our findings.

In the spin glass phase, the system consists of macroscopic spin
domains of variable sizes.
Each of these domains flips as a coherent
entity, and the flipping costs only a finite free energy.
The variability in size gives rise to a hierarchical structure in
state space. At each level in the hierarchy
some state clusters split; each such splitting is associated with
a spin-domain. The first (highest) level splitting (to $\C, \bar
\C$) is associated with the largest domain $\G_1$; at the next
level the two observed splittings ($\C \rightarrow \C_1,\C_2$ and
$\bar \C \rightarrow \bar \C_1,\bar \C_2$) are related by
symmetry and hence governed by the same, second largest domain
$\G_2$. At each level, the state clusters are labeled according to
the orientation of the corresponding domains.

Below the second level,
different spin domains are involved depending on which state cluster
is being subdivided,
e.g. $\G_3$ is the domain whose orientation splits the states in
$\C_1$, while a different domain $\G_3^\prime$ is involved in
splitting $\C_2$. Although $\G_3\not=\G_3^\prime$, in general they
may share some of their spins. The state space structure in the
lower levels of the hierarchy has to be further investigated for
larger systems. Specifically, one has to verify that $\G_3$ and
$\G_3^\prime$ do not vanish as $L \rightarrow \infty$ .

Some details of our hierarchical picture 
do not appear to be
consistent
with RSB. According to the RSB scenario, the states have an
ultrametric structure, which implies that for any two
state clusters defined at a certain level of the hierarchy, e.g.
$\C_1$ and $\C_2$, the distribution of overlaps $q_{ij}$ between
$i\in\C_1$ and $j\in\C_2$ should approach a delta-function for
large $L$.  We presented in Sec. \ref{sec:states} indications that
the width of the distribution $P(\tilde D_{ij})$, of values in
$\tilde D$,
may not vanish for $L \rightarrow \infty$,
indicating absence of ultrametricity.
We also presented direct evidence
for lack of ultrametricity
in Sec.~\ref{sec:ultm}.
However, studies on
larger sizes are needed to verify that
the test which indicates lack of ultrametricity will still yield the same
conclusion
as $L \rightarrow \infty$.

In Sections \ref{sec:overlap} and \ref{sec:ultm} we demonstrated
how, by
separating the state space into its components,
we can calculate various quantities using
only a chosen part of this space, thus obtaining more
reliable numerical results and reducing finite size effects.

Clustering analysis can be applied also to other systems with a
non-trivial phase space structure, {\em i.e.}\/ which have several
valleys which are not related by any apparent symmetry, such
as random field models, see {\em e.g.}\/ the discussion in
Ref.~\onlinecite{Huse87}, or other models with random anisotropy
\cite{Fisch95}. It can help not only in the investigation of the
macroscopic properties of a system, but also in understanding the
micro-structure that give rise to its properties.

{ \bf Acknowledgements} The research of ED was partially supported by 
grants from the Germany-Israel Science Foundation (GIF);
his stay at the ITP, UCSB was supported in part by the NSF under grant PHY99-07949.
APY acknowledges
support from the NSF through grant DMR 0086287.  
We thank A. Hartmann,
I. Kanter, E. Marinari and M. M\'ezard for most helpful discussions.

\bibliography{spngls18}

\begin{thebibliography}{10}
\bibitem{foot2}
Strictly, it is not an equilibrium canonical ensemble, since for each state in
  the ensemble we have also its exact inverse, but this does not affect our
  results.
\bibitem{foot7}
If say $\G_1$, the spin domain that flips, consists of two disjoint parts
  $\G_1^a$ and $\G_1^b$, one can flip each of these separately, one at a time.
  In such a case we identify the largest of the two sub-domains as our $\G_1$,
  whereas the smaller induces the finer level splitting in the state hierarchy.
\bibitem{foot3}
Except maybe a small part of $\G_a$, that may lose its relative orientation due
  to local thermal excitations.
\bibitem{foot4}
The leaves of the hierarchy discussed here are {\it not} the individual states
  {\bf S}$^\mu$ of the state dendrogram of Sec. \ref{sec:states}.
\bibitem{foot6}
Ultrametricity is trivially violated if we include in the same triangle states
  with negative and positive overlaps.
\end{thebibliography}
\bibliographystyle{prsty}

\end{document}